\documentclass{raa}

\usepackage{graphicx,times}
\usepackage{natbib}
\usepackage{amssymb,amsmath}
\usepackage{longtable}
\bibpunct{(}{)}{;}{a}{}{,}

\usepackage[pagebackref=true]{hyperref}

\begin{document}

\title{Red Quasars: Selecting Candidates in SDSS DR16 and Estimating Their Physical Parameters}

\volnopage{ {\bf 20XX} Vol.\ {\bf X} No. {\bf XX}, 000--000}
  \setcounter{page}{1}

\author{M.Yu. Piotrovich\inst{}, S.D. Buliga\inst{}, T.M. Natsvlishvili\inst{}}

\institute{Central astronomical observatory at Pulkovo, St.-Petersburg, 196140, Russia;
{\it mpiotrovich@mail.ru}\\
\vs \no
   {\small Accepted for publication if Research in Astronomy and Astrophysics.}
}

\abstract{Using ''color cut'' method we obtained from SDSS DR16 catalog 733 red quasar candidates which amounted to approximately 4\% of the objects from the initial sample. Then we estimated the radiative efficiency, spins, inclination angles, and corresponding new SMBH masses for all 733 objects using three theoretical models. Obtained spin distributions contain a large percentage of objects with retrograde rotation. It may indicate that these are either very young objects or objects that formed as a result of mergers. The dependencies of the estimated spin values on SMBH masses show strong correlation with linear fit slope 0.9-1.0 which allows us to assume that red quasars are likely to contain both Seyferts and NLS1, and that the main mechanism of SMBH mass growth in these objects is disk accretion.
}

   \authorrunning{M.Yu. Piotrovich et al.}
   \titlerunning{Red Quasars: Selecting Candidates in SDSS DR16 and Estimating Their Physical Parameters}
   \maketitle

\section{Introduction}

Quasars represent one of the most luminous manifestations of accretion onto supermassive black holes (SMBHs) and play a fundamental role in the co-evolution of black holes and their host galaxies \citep{salpeter64,lyndenbell69,kormendy13}. While optically selected quasar samples have been extensively studied for several decades \citep{schmidt63,richards06}, they are inherently biased against dust-obscured systems. As a result, a significant fraction of the quasar population may be
missed in traditional optical surveys.

Red quasars constitute a distinct class of active galactic nuclei (AGN) characterized by unusually red optical and near-infrared colors \citep{webster95,glikman04}. Their redness is commonly attributed to dust extinction along the line of sight \citep{hopkins05,glikman07}, although intrinsic spectral properties and host-galaxy contamination may also contribute \citep{urry95,calderone12}. Red quasars are often absent from optical color-selected samples but can be efficiently identified through infrared and radio surveys \citep{cutri01,white03}. As such, they provide a more complete view of the AGN population and probe phases of black hole growth that are otherwise obscured.

An increasing body of observational evidence suggests that red quasars may represent a transitional stage in quasar and galaxy evolution \citep{sanders88,hopkins08}. In this scenario, rapid SMBH accretion occurs while the central engine is still embedded in a dust-rich environment, possibly triggered by major galaxy mergers \citep{treister12,glikman15}. Feedback from the AGN is expected to eventually expel the surrounding gas and dust, revealing an unobscured blue quasar \citep{fabian12}. This evolutionary picture is supported by the high Eddington ratios, disturbed host morphologies, and powerful outflows observed in many red quasar samples \citep{urrutia08,banerji12,brusa15}.

Despite their importance, the physical properties of red quasars remain less well constrained than those of unobscured quasars.

In our previous work \citep{piotrovich24} we studied a sample of 42 reliably identified red quasars from the literature. In this work, in order to increase the number of objects studied, we decided to search for the candidates for red quasars ourselves in publicly available catalogs. We took all the data from the literature, but it should be noted that uncertainties in dust extinction, inclination effects, and host-galaxy contribution complicate the determination of fundamental parameters such as bolometric luminosity, SMBH mass, and accretion efficiency \citep{vasudevan09,calderone13} and this problem undoubtedly requires further study.

\section{Obtaining and analysis of the red quasar sample}

\begin{figure}[ht!]
\includegraphics[bb= 30 10 715 530, clip, width=0.5\linewidth]{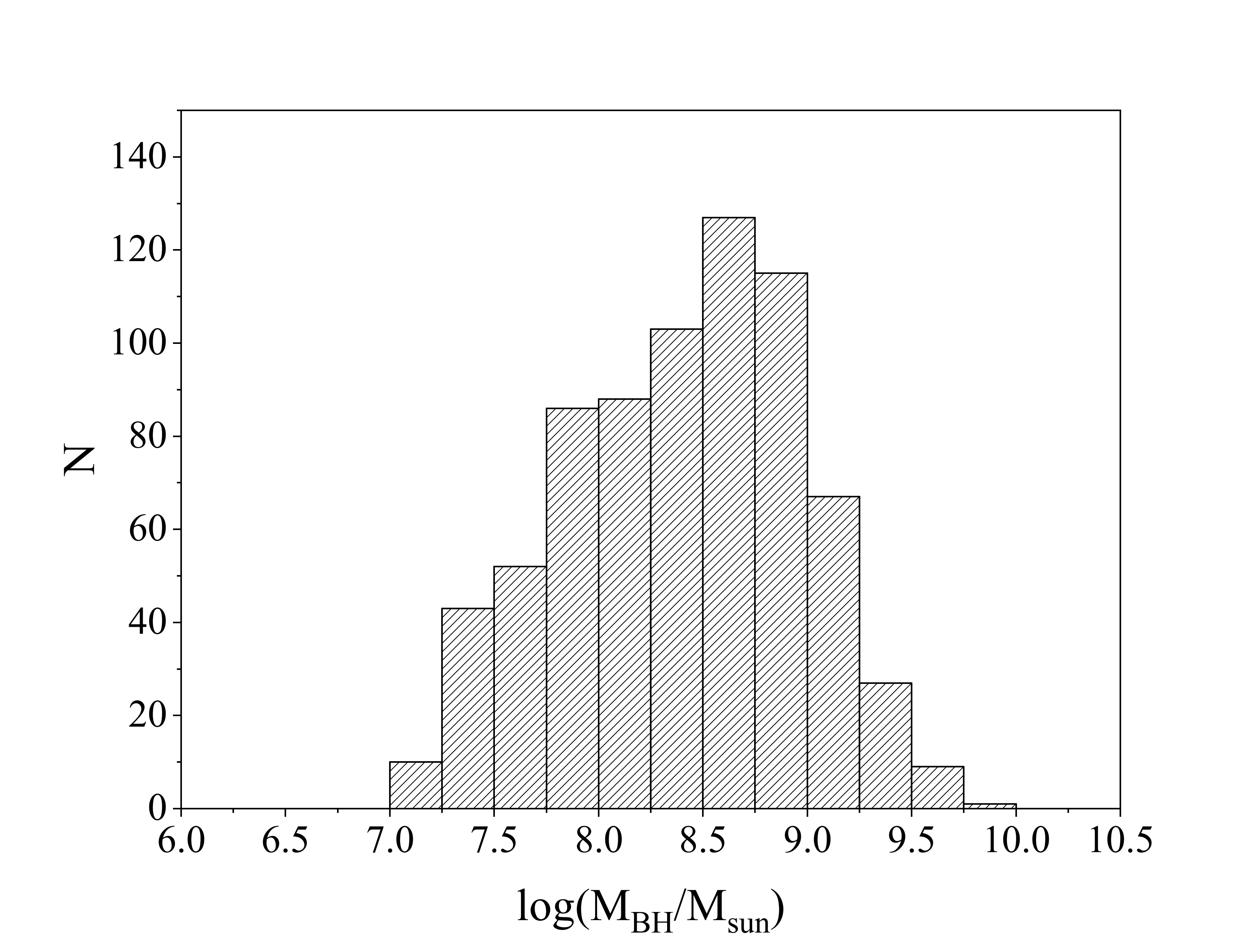}
\includegraphics[bb= 30 10 715 530, clip, width=0.5\linewidth]{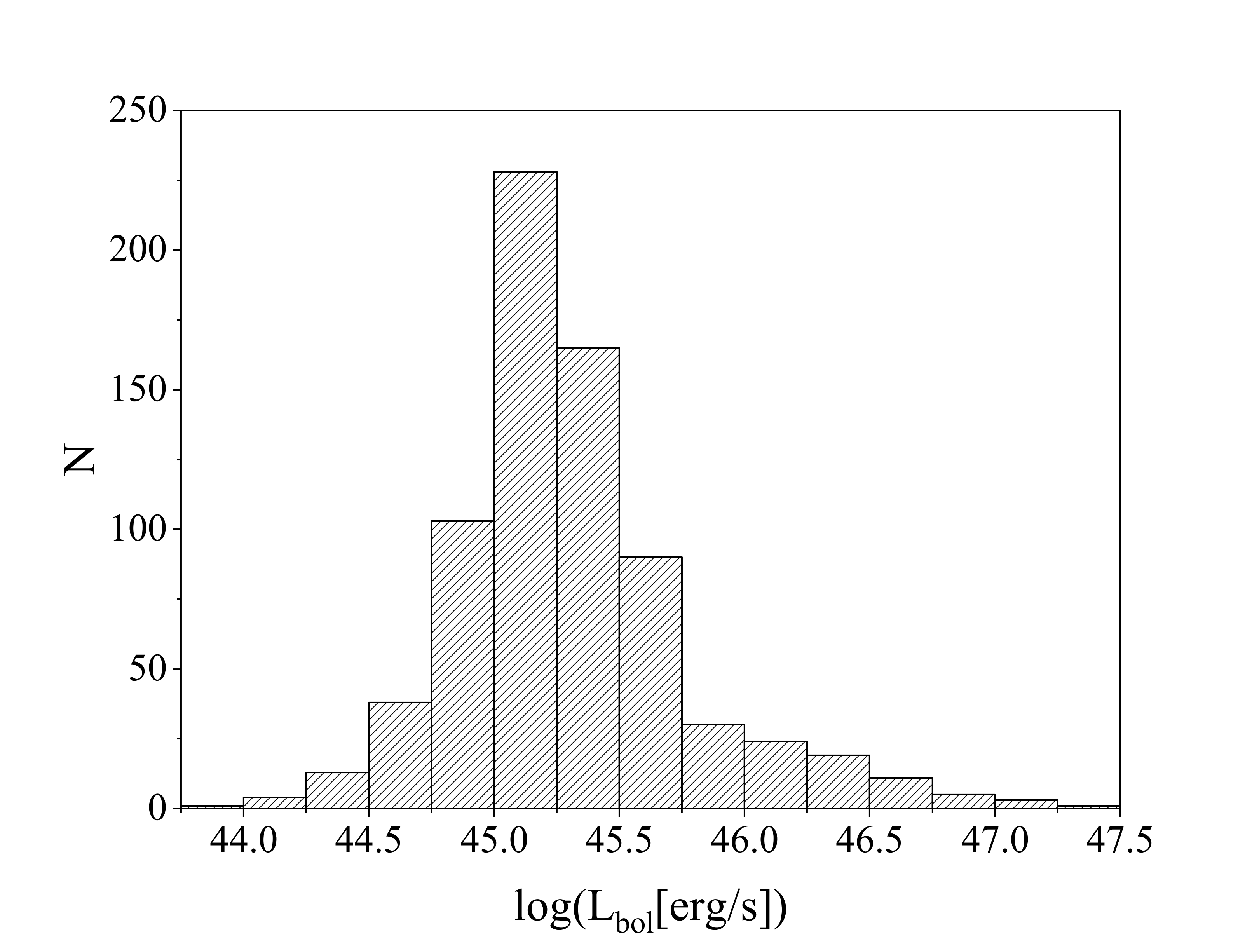}
\includegraphics[bb= 30 10 715 530, clip, width=0.5\linewidth]{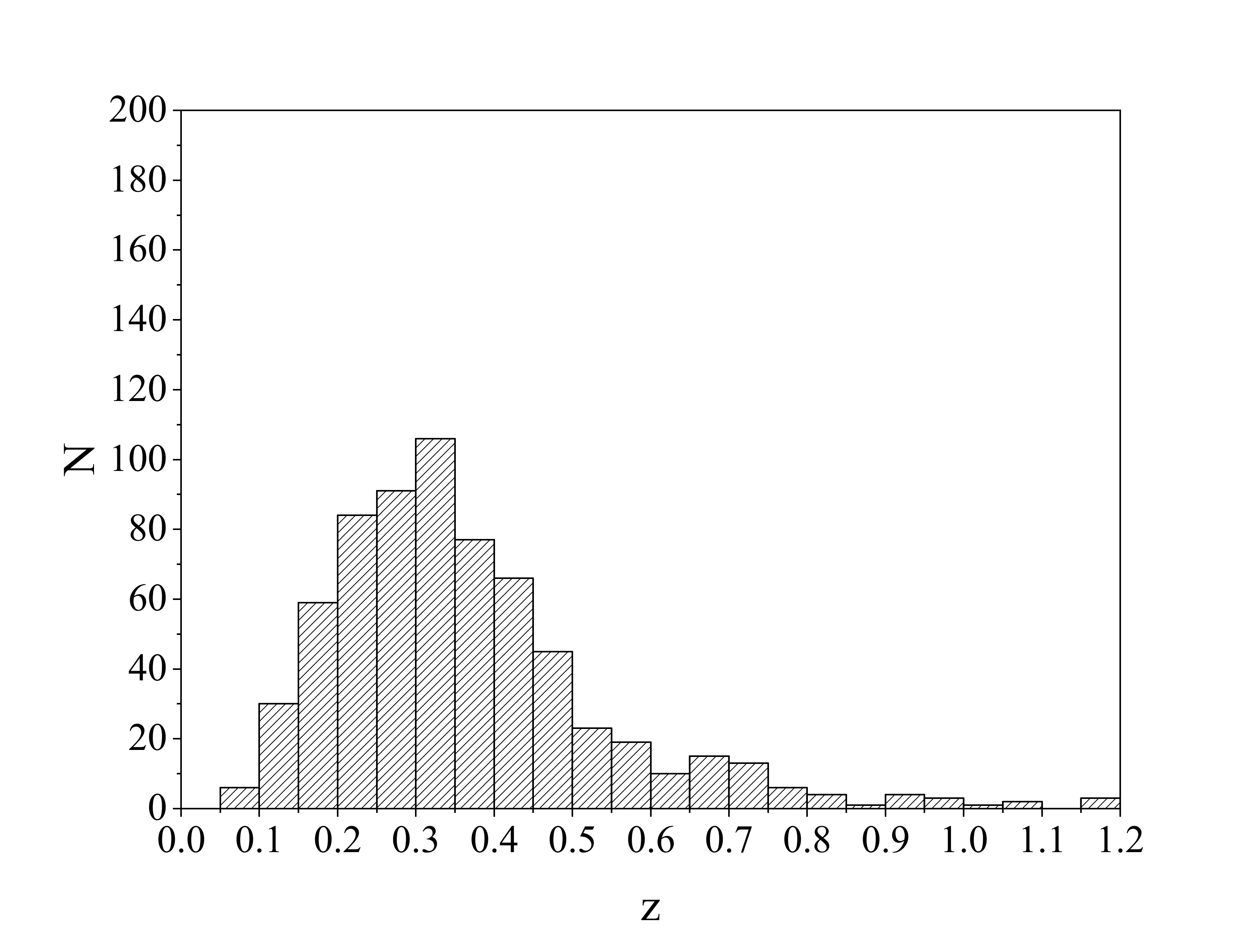}
\includegraphics[bb= 30 10 715 530, clip, width=0.5\linewidth]{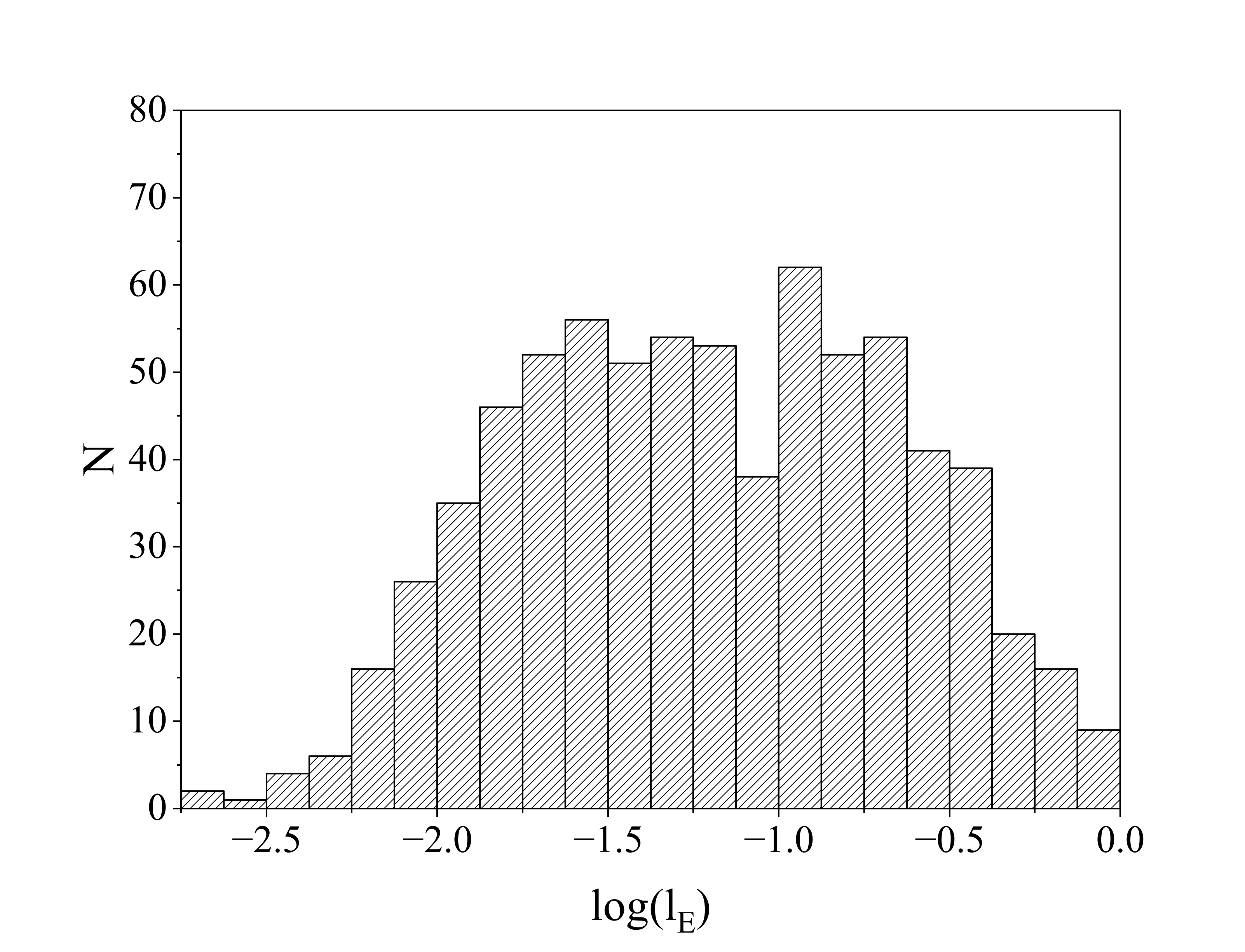}
\caption{Distribution of the SMBH mass, the bolometric luminosity, the cosmological redshift and the Eddington ratio in our red quasar sample.
\label{fig:hist_MBH_Lbol_z_lE}}
\end{figure}

\begin{figure}[ht!]
\includegraphics[bb= 30 10 715 535, clip, width=0.5\linewidth]{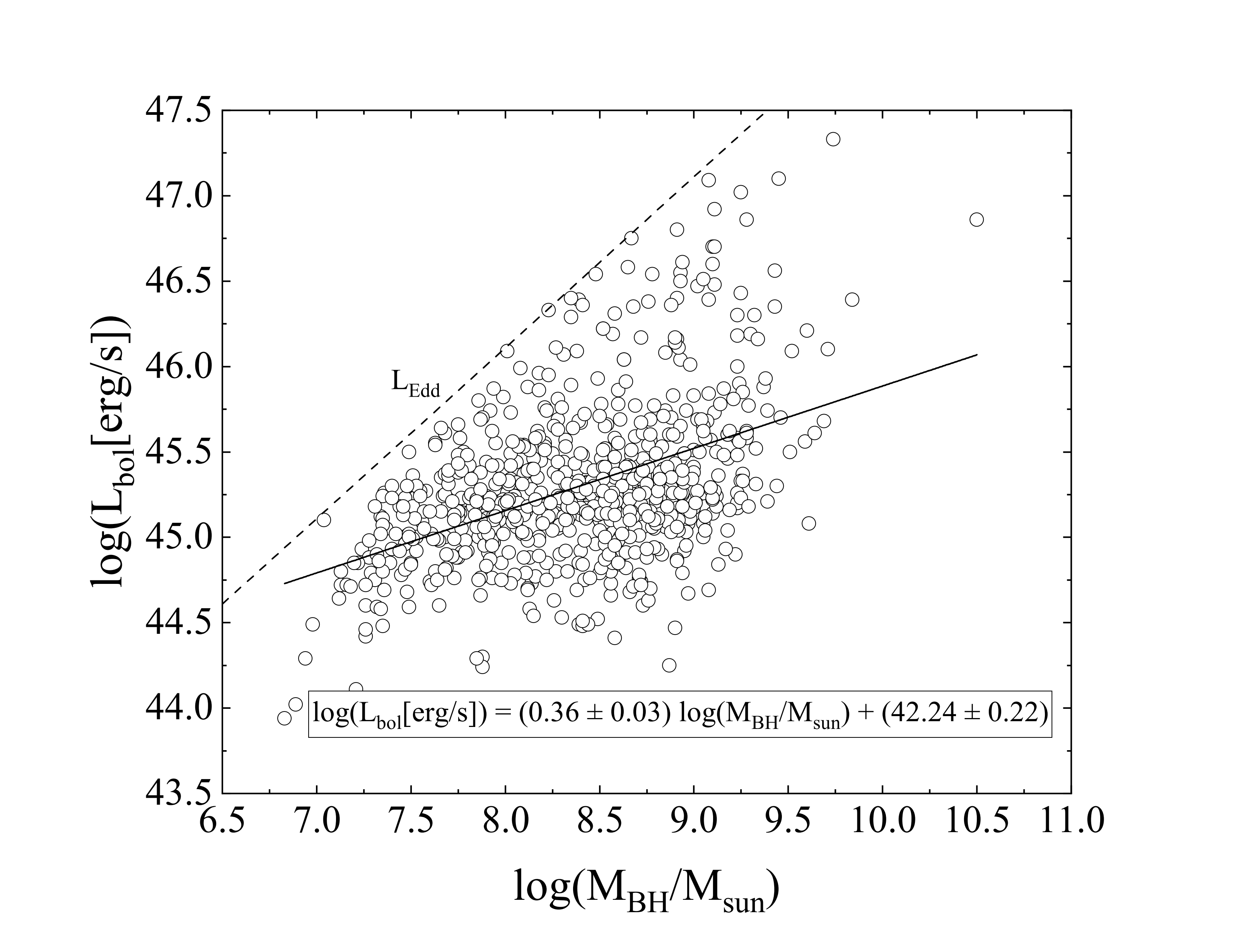}
\includegraphics[bb= 30 10 715 535, clip, width=0.5\linewidth]{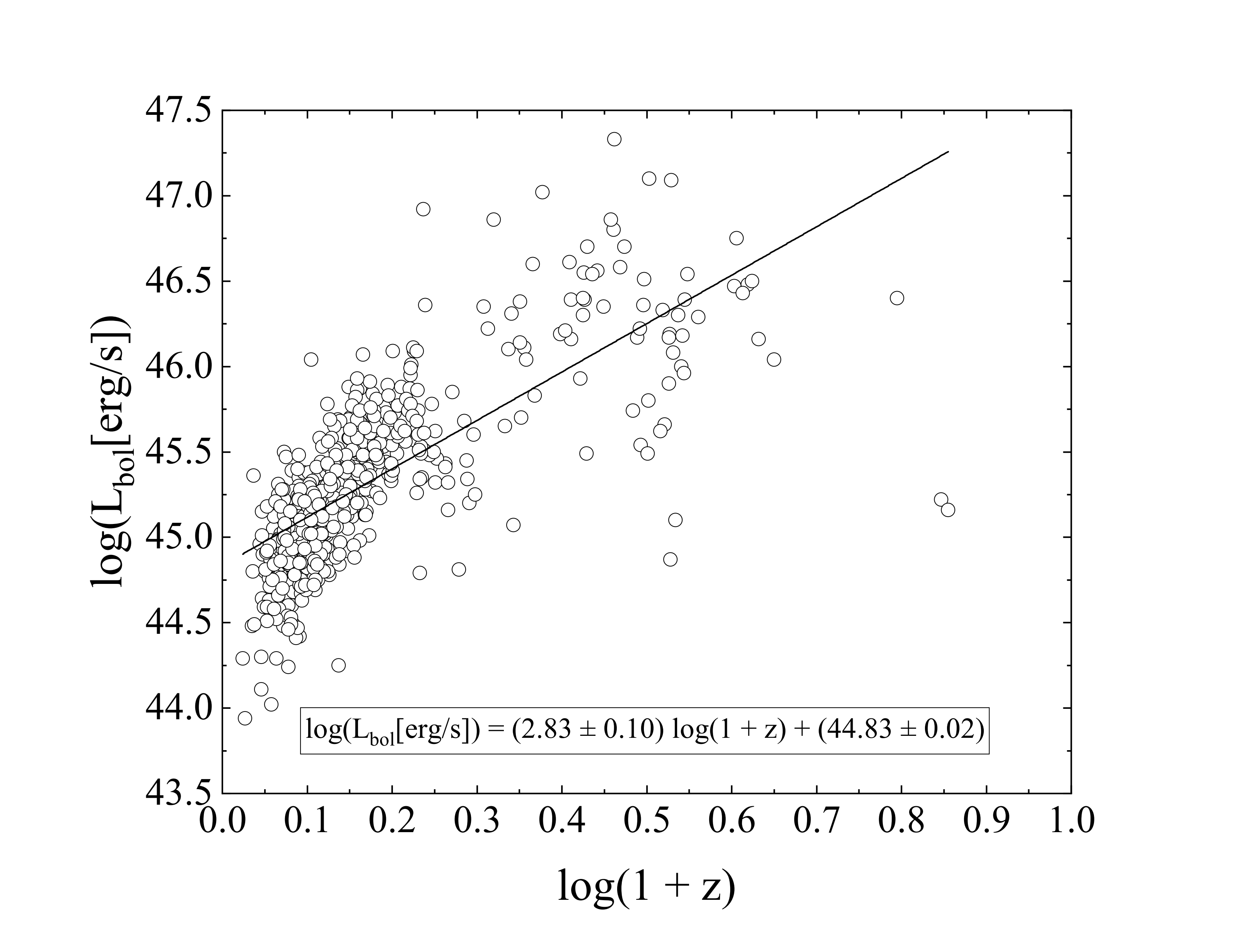}
\caption{Dependencies of the bolometric luminosity on the SMBH mass and the cosmological redshift in our initial red quasar sample. $\log{L_{\rm Edd}{\rm [erg/s]}} = \log(M_{\rm BH} / M_\odot) + 38.11$ is the Eddington luminosity.
\label{fig:Lbol_MBH_z}}
\end{figure}

\begin{figure}[ht!]
\includegraphics[bb= 30 10 715 545, clip, width=\linewidth]{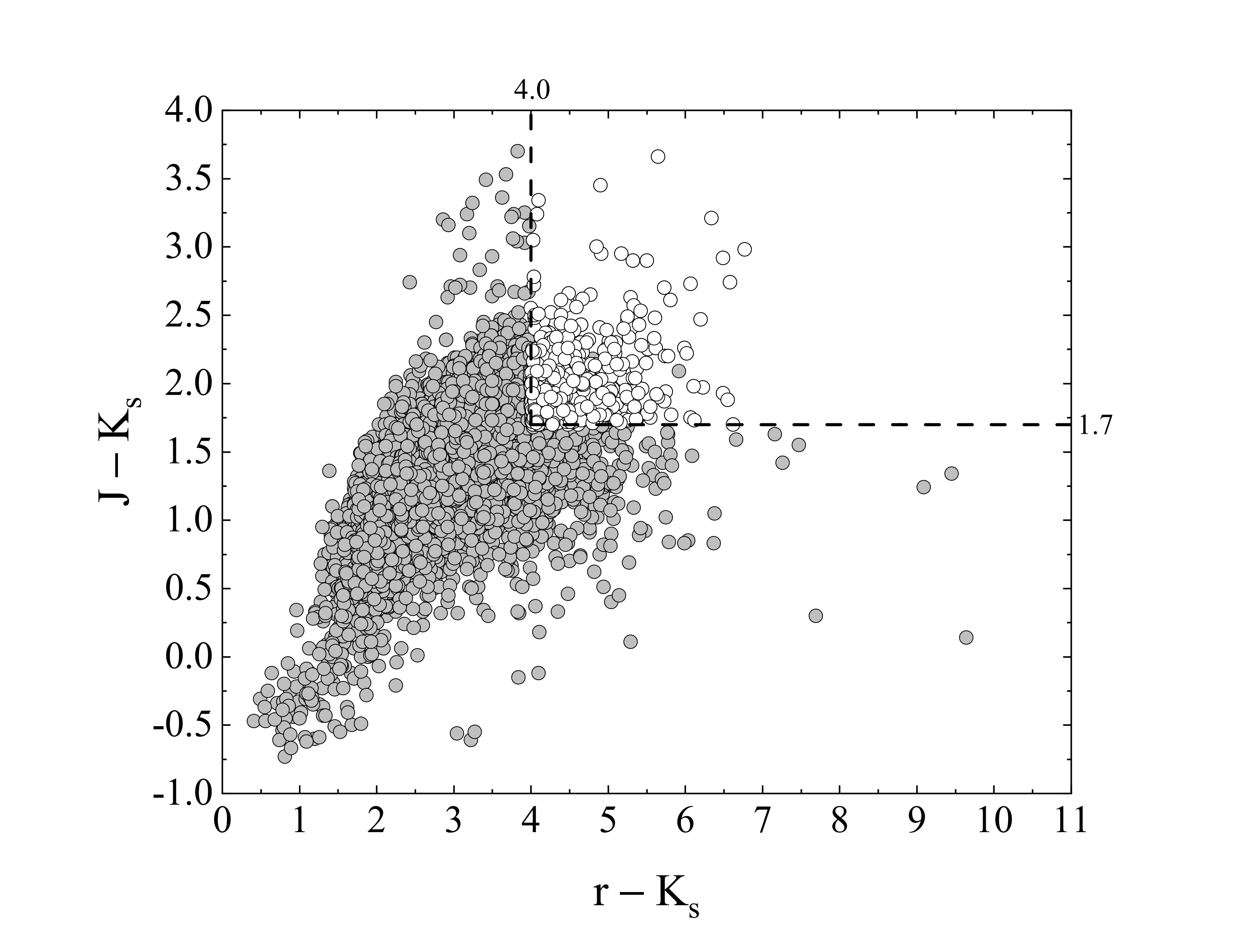}
\includegraphics[bb= 30 10 715 535, clip, width=0.5\linewidth]{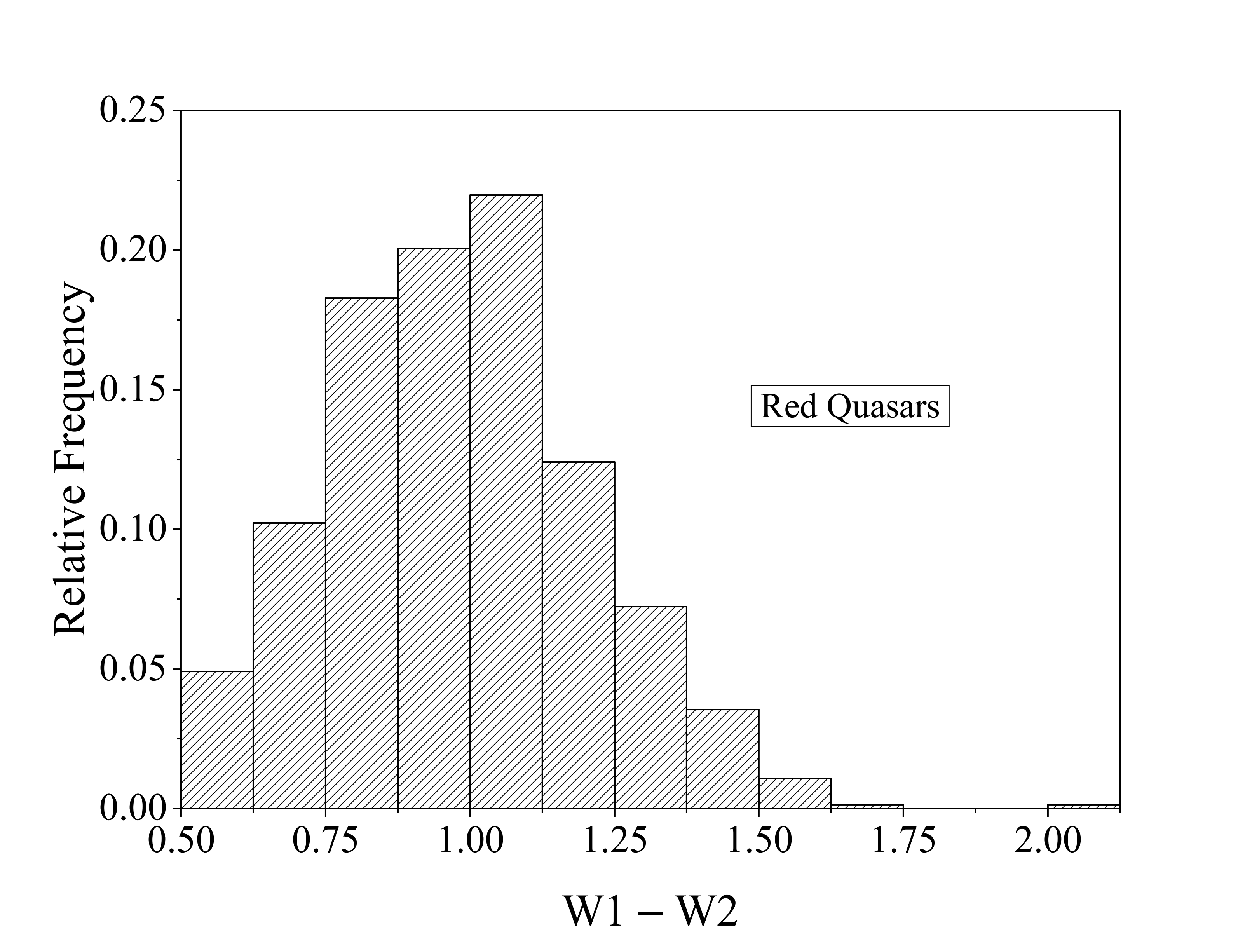}
\includegraphics[bb= 30 10 715 535, clip, width=0.5\linewidth]{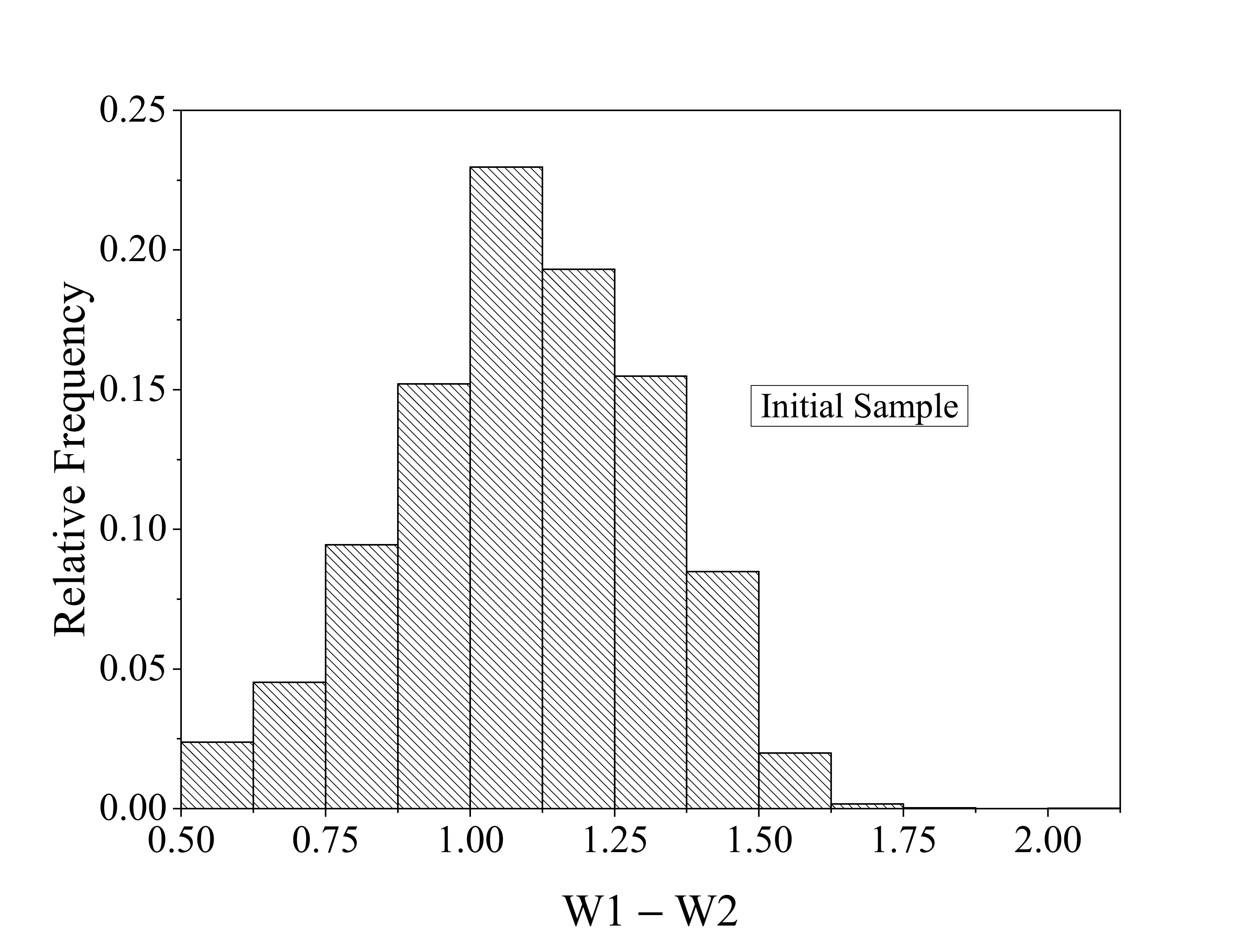}
\caption{Colors of objects of our initial sample expressed in $(J - K_{\rm s})$ and $(r - K_{\rm s})$ color parameters (the red quasars candidates we selected are marked with white circles) and distribution of (W1 - W2) color parameter for the initial sample and for the red quasars.
\label{fig:colors}}
\end{figure}

As a data source, we used the catalog from \citet{wu22}, because it contains all the data necessary for our estimations. It consists of continuum and emission-line properties for 750414 broad-line quasars included in the Sloan Digital Sky Survey Data Release 16 quasar catalog (SDSS DR16Q).

In our paper, we use the same cosmological model used in \citet{wu22} catalog: a flat ${\rm \Lambda}$CDM cosmology with $\Omega_{\rm \Lambda} = 0.7$, $\Omega_M = 0.3$, and $H_0 = 70$km s$^{-1}$ Mpc$^{-1}$.

Also in this work we use a methodology similar to that of \citet{glikman12}: data on magnitudes from the SDSS, which are originally in the AB units, were converted into Vega units, and data from the WISE and 2MASS are used in the Vega units.

The criteria for red quasars can be the difference in the magnitudes in the bands: $J - K_{\rm s} > 1.7$ and $r - K_{\rm s} > 4.0$ \citep{glikman07,glikman12}. In order to calculate $r – K_{\rm s}$ we converted $r$ from AB to Vega units: $r_{\rm Vega} = r_{\rm AB} – 0.226$. Additionally, the difference between bands $W1$ and $W2$ can also be used: $W1 - W2 > 0.5$ \citep{wright10,stern12,richards15,glikman22}. It should be noted that this ''color cut'' method is not 100\% reliable, so our objects should be considered as candidates for red quasars.

Of all the objects in the catalog, 17926 objects have all the data needed for our calculations. Using conditions $J - K_{\rm s} > 1.7$, $r - K_{\rm s} > 4.0$ and $W1 - W2 > 0.5$, we selected 744 objects from the catalog. Two objects have a significantly higher value of the color parameter $r - K_{\rm s}$. These are SDSS~094602.11+003518.5 ($r - K_{\rm s} = 9.43,\, z \approx 0.649$) and SDSS~095932.74+452330.5 ($r - K_{\rm s} = 9.94,\, z \approx 1.690$). Analysis revealed that these two objects had serious issues with their type identification and measured spectra. Therefore, we decided to remove them from our sample. Thus, about 4\% of the objects from the studied sample turn out to be candidates for red quasars. In \citet{kim24} the authors found that red Type-1 quasars make up approximately 10\% of the total number of Type-1 quasars. The lower percentage in our case may be explained by the fact that our sample does not contain only Type-1 quasars, and also because our method does not detect all red quasars.

Since our spin estimation method assumes standard Shakura--Sunyaev accretion disk model \citep{shakura73}, we additionally removed from the sample 9 objects whose Eddington ratio ($l_{\rm E} = L_{\rm bol} / L_{\rm Edd}$) was greater than 1.

We then analyzed the statistical properties of the 733 red quasar candidates obtained.

Fig.\ref{fig:hist_MBH_Lbol_z_lE} shows distribution of the SMBH mass $M_{\rm BH}$, the bolometric luminosity $L_{\rm bol}$, the cosmological redshift $z$ and the Eddington ratio $l_{\rm E}$ in our red quasar sample. It can be seen that all distributions are close to log-normal (for $z$ - normal) shape and in general have an appearance characteristic of most quasars. Note that the bolometric luminosities in \citet{wu22} were obtained using bolometric corrections calculated for a wide range of AGNs, and this method can produce significant inaccuracies specifically for red quasars. However, due to the relatively limited study of red quasars at present, we believe that a more accurate determination of bolometric luminosities is currently an extremely difficult task and certainly requires further study.

Fig.\ref{fig:Lbol_MBH_z} shows dependencies of the bolometric luminosity on the SMBH mass and the cosmological redshift in our red quasar sample. First dependence has moderate correlation between $L_{\rm bol}$ and $M_{\rm BH}$ (Pearson correlation coefficient is 0.45). Second dependence has strong correlation between $L_{\rm bol}$ and $z$ (Pearson correlation coefficient is 0.73). Linear fitting gives us:
\begin{equation}
  \begin{aligned}
    &\log(L_{\rm bol}{\rm [erg/s]}) = (0.36 \pm 0.03) \log(M_{\rm BH}/M_\odot) + (42.24 \pm 0.22),\\
    &\log(L_{\rm bol}{\rm [erg/s]}) = (2.83 \pm 0.10) \log(1 + z) + (44.83 \pm 0.02)
  \end{aligned}
\end{equation}

The linear fit slope $0.36 \pm 0.03$ of the bolometric luminosity on the SMBH mass dependence is noticeably less than similar values obtained by us in previous works. For example, for small sample of red quasars we had $0.88 \pm 0.12$ value \citep{piotrovich24}, for local active galactic nuclei (AGNs) $1.19 \pm 0.08$ \citep{piotrovich22}, for distant low luminosity AGNs (LLAGNs) $0.73 \pm 0.09$ \citep{piotrovich25b}. It should also be noted that in all these cases the correlation coefficient was significantly higher. Also, if you look at Fig.2 from \citet{wu22}, you can see that for all objects in the catalog as a whole, the slope appears to be close to 1. Apparently, this feature of red quasars is related to the fact that at bolometric luminosities $\log(L_{\rm bol}{\rm [erg/s]}) \gtrsim 46$, the number of red quasars decreases sharply. It can be assumed that at high luminosities dust (which, according to modern concepts, causes the reddening of the spectrum) evaporates and/or is blown out by the light pressure.

The strong dependence of the bolometric luminosity on the redshift is most likely caused by the selection bias, because at a great distances we observe mostly the brightest objects.

In Fig.\ref{fig:colors} one can see the colors of the objects of our initial sample (of 17926 objects) expressed in ($J - K_{\rm s}$) and ($r - K_{\rm s}$) color parameters (the red quasars candidates we selected are marked with white circles) and the distribution of ($W1 - W2$) color parameter for the initial sample and for the red quasars. The top figure can be compared with a similar figure from \citet{glikman12} (Fig.1). In general, our objects are located in the same area as in \citet{glikman12}.

The distribution of the ($W1 - W2$) color parameter for the red quasar candidates are close to the log-normal shape with the peak at 1. It looks similar to the analogous distribution of red quasars at Fig.2 from \citet{glikman22}. The analogous distribution for the initial sample looks similar, with the peak in the same location, but the distribution pattern to the left and right of the peak is slightly different, making it look even more log-normal, which is most likely due to the fact that there are $\sim$25 times more objects here and they are of different types.

\section{Method for estimating physical parameters of SMBHs and accretion disks}

In this work we, for self-consistency, used method similar to our previous works \citep{piotrovich22,piotrovich24,piotrovich25b}. It should be noted that our method is highly model-dependent. However, less model-dependent methods using such a set of initial data currently do not exist and the indirect confirmation of the validity of this method is the fact that it yields significantly different results for different types of AGN and quasars which lend themselves to a plausible phenomenological explanation in terms of different physical processes in these objects \citep{piotrovich22,piotrovich23,piotrovich25,piotrovich25b}.

\begin{figure}[ht!]
\centering
\includegraphics[bb= 30 10 715 535, clip, width=0.7\linewidth]{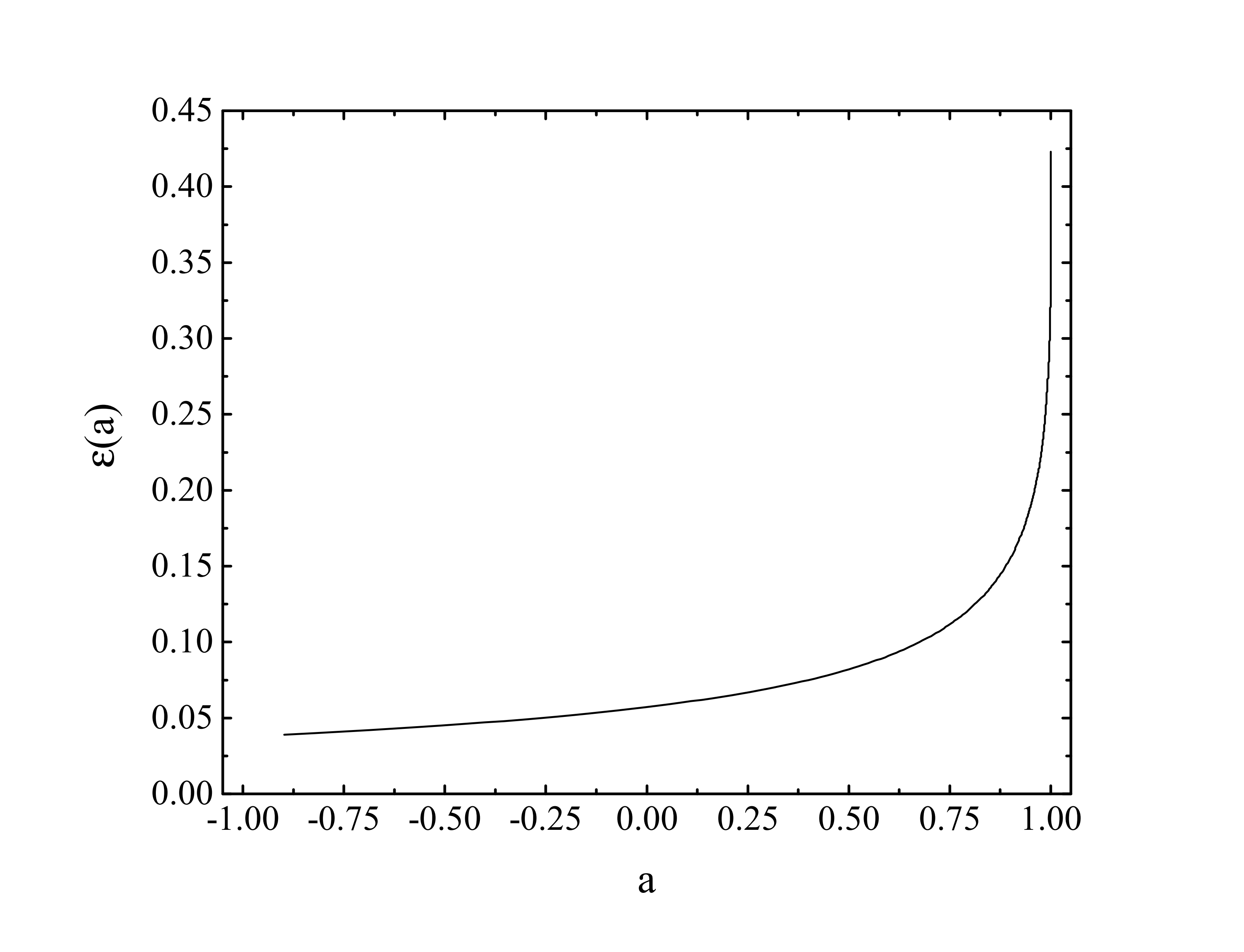}
\caption{Dependence of the radiative efficiency on the BH spin.
\label{fig:epsilon}}
\end{figure}

The spin of a black hole can be estimated via the radiative efficiency of its accretion disk (see Fig.\ref{fig:epsilon}), $\varepsilon(a)$, defined as
\begin{equation}
  \varepsilon = \frac{L_{\rm bol}}{\dot{M}c^2},
\end{equation}
where $\dot{M}$ is the accretion rate, $0.039 < \varepsilon < 0.324$ and $-1 \le a \le 0.998$ \citep{thorne74}. Negative values of $a$ correspond to retrograde rotation.

Radiative efficiency is related to observable AGN parameters within the framework of the Shakura--Sunyaev thin-disk model \citep{shakura73}. In this work, three statistical models are adopted:
\begin{enumerate}
\item \citet{du14}:
\[
\varepsilon = 0.105
\left(\frac{L_{\rm bol}}{10^{46}}\right)
\left(\frac{L_{5100}}{10^{45}}\right)^{-1.5}
M_8 \mu^{1.5};
\]

\item \citet{raimundo11}:
\[
\varepsilon = 0.063
\left(\frac{L_{\rm bol}}{10^{46}}\right)^{0.99}
\left(\frac{L_{\rm opt}}{10^{45}}\right)^{-1.5}
M_8^{0.89} \mu^{1.5};
\]

\item \citet{trakhtenbrot14}:
\[
\varepsilon = 0.073
\left(\frac{L_{\rm bol}}{10^{46}}\right)
\left(\frac{\lambda L_\lambda}{10^{45}}\right)^{-1.5}
\left(\frac{\lambda}{5100\,\text{\AA}}\right)^{-2}
M_8 \mu^{1.5}.
\]
\end{enumerate}

Here $M_8 = M_{\rm BH}/10^8M_\odot$, $\mu=\cos i$ and $i$ is an inclination angle between the line of sight and the normal to the accretion disk plane. When $L_{5100}$ measurements were unavailable, we adopted the bolometric correction $L_{5100}=L_{\rm bol}/10.3$ \citep{richards06}. The inclination angle was initially set to $i=45^\circ$ and then the angle value was changed up or down in 5 degree increments until model we were using produced a physically meaningful result.

We chose these models because they are sufficiently different from each other to be considered distinct models rather than variants of a single model. We use these models here for greater consistency with many of our previous works where we have also used them. The problem with choosing a single model is that each one describes its own type of object slightly better. However, we don't know exactly which type of AGN predominates among the red quasars.

The standard method (which was used by \citet{wu22}) for determining the mass of a SMBH from broad line region parameters is as follows:
\begin{equation}
M_{\rm BH}=\frac{R_{\rm BLR}V_{\rm BLR}^2}{G}, \qquad
V_{\rm BLR}\simeq \frac{\mathrm{FWHM(H}\beta)}{2\sin i},
\end{equation}
where $R_{\rm BLR}$ is broad line region radius and $V_{\rm BLR}$ is velocity of accreting matter at that radius.

Note that \citet{wu22} determined the masses using different spectral lines for objects at different redshifts (see paragraph 4.1 of their paper) using the standard assumption that $i \approx 30^\circ$. The method described above uses the FWHM(H$\beta$) as an example, which is only applicable for relatively nearby objects.

To take into account the dependence of SMBH mass on $i$, we recalculated the mass value. Note that our mass conversion method is actually independent of the spectral line used to determine the initial mass. The mass depends on inclination angle as follows: $M \sim 1 / (2 \sin{(i)})^2$. This is a purely geometric effect related to the orientation of the accretion disk relative to the observer. It is in no way dependent on the specific spectral line used to estimate the mass. Therefore, if we take a different angle $i_{\rm new}$, the new mass will be:
\begin{equation}
M_{\rm new} = M (\sin{(i)} / \sin{(i_{\rm new})})^2,
\end{equation}
regardless of which line was used to determine the mass.

The spin was then determined numerically by solving the $\varepsilon(a)$ relation for the ISCO radius \citep{bardeen72}:
\begin{equation}
\varepsilon(a)=1-\frac{R_{\rm ISCO}^{3/2}-2R_{\rm ISCO}^{1/2}+|a|}
{R_{\rm ISCO}^{3/4}(R_{\rm ISCO}^{3/2}-3R_{\rm ISCO}^{1/2}+2|a|)^{1/2}},
\end{equation}
where $R_{\rm ISCO}$ is the radius of the innermost stable circular orbit:
\begin{equation}
  \begin{array}{l}
   R_{\rm ISCO}(a) = 3 + Z_2 \pm [(3 - Z_1)(3 + Z_1 + 2 Z_2)]^{1/2},\\
   Z_1 = 1 + (1 - a^2)^{1/3}\left[(1 + a)^{1/3} + (1 - a)^{1/3}\right],\\
   Z_2 = (3 a^2 + Z_1^2)^{1/2}.
  \end{array}
\end{equation}
In this expression ''-'' is used when $a \geq 0$, and ''+'' when $a < 0$.

\section{Analysis of the estimated parameters of the red quasars candidates}

\begin{figure}[ht!]
\includegraphics[bb= 30 5 715 535, clip, width=0.5\linewidth]{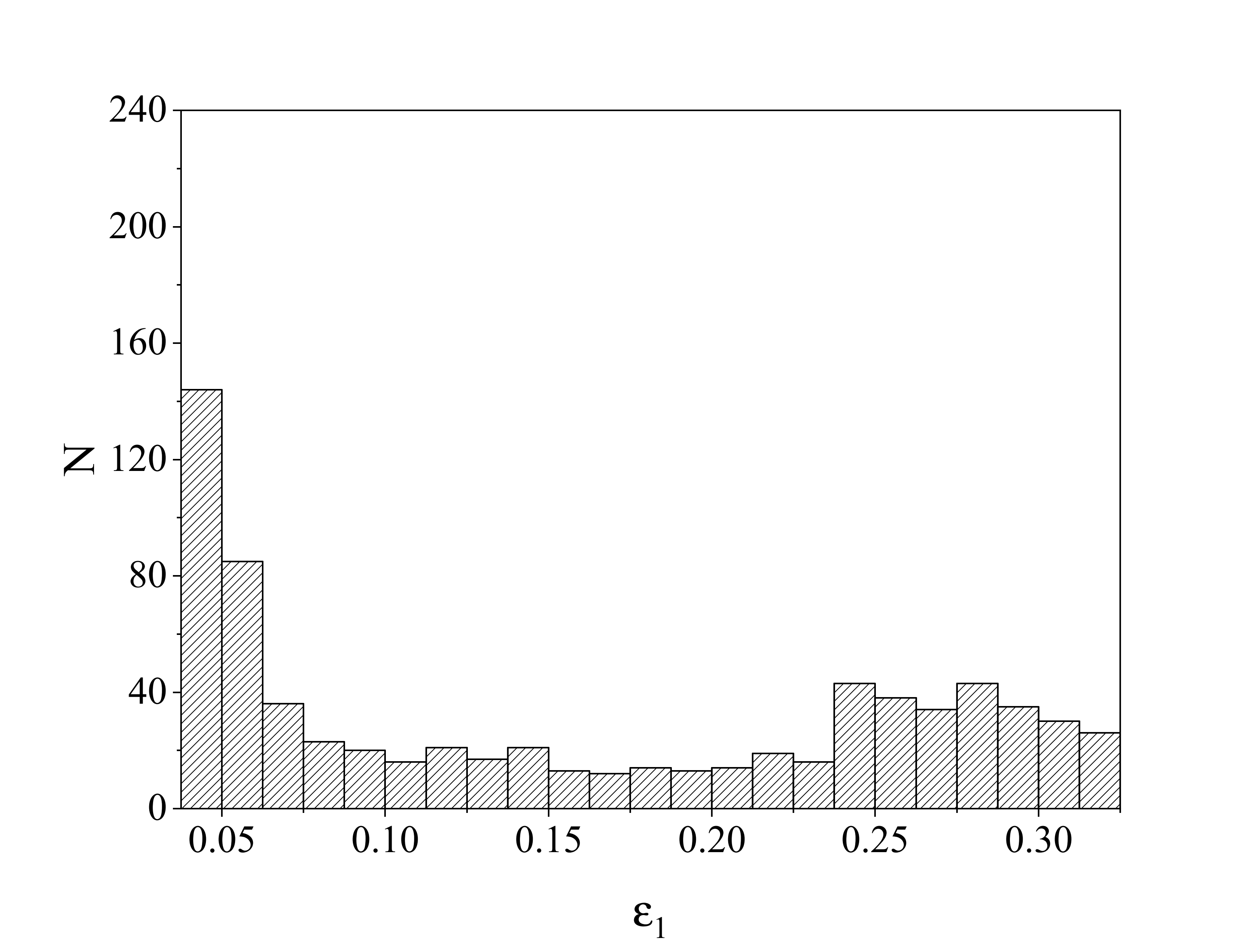}
\includegraphics[bb= 30 10 715 535, clip, width=0.5\linewidth]{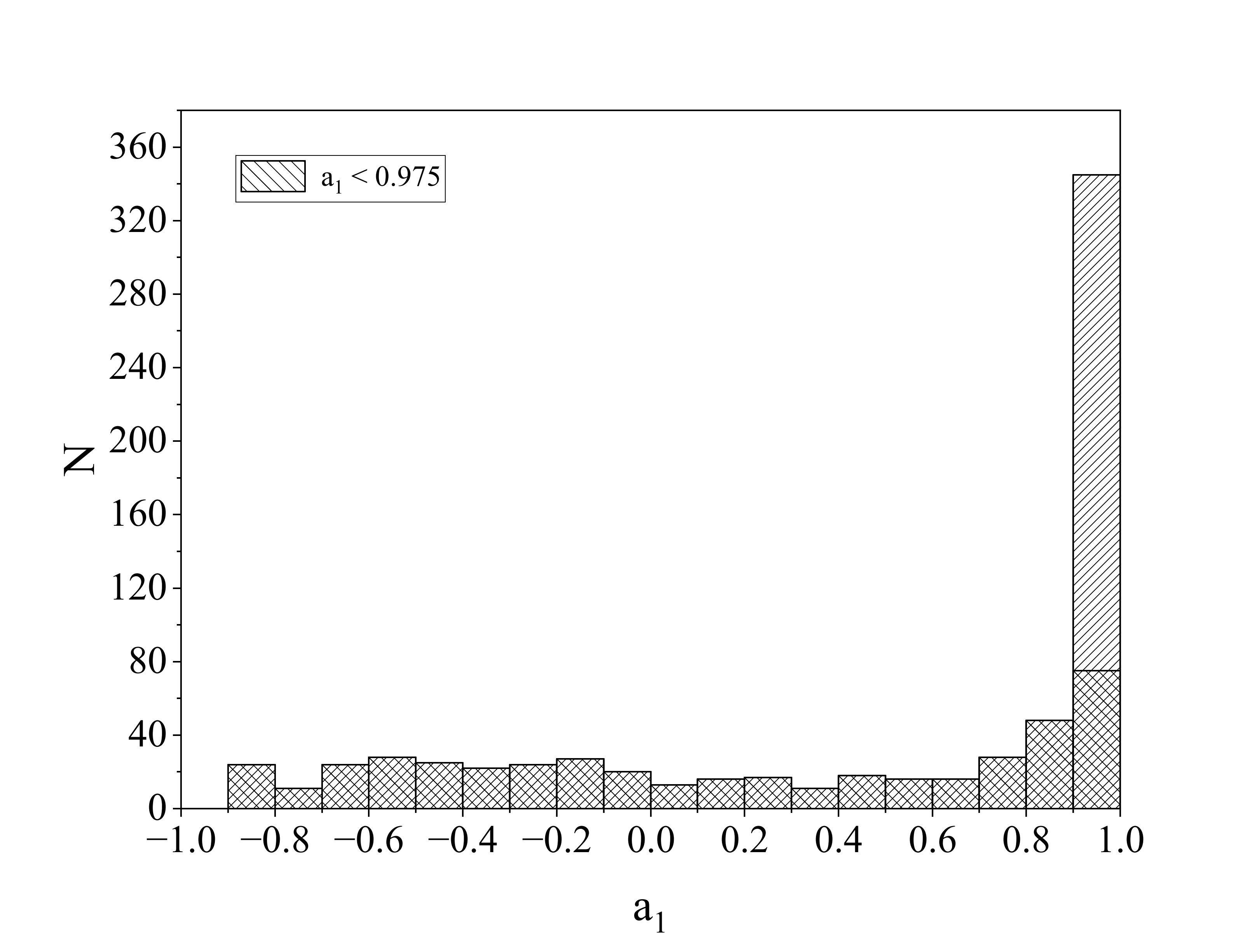}
\includegraphics[bb= 30 5 715 535, clip, width=0.5\linewidth]{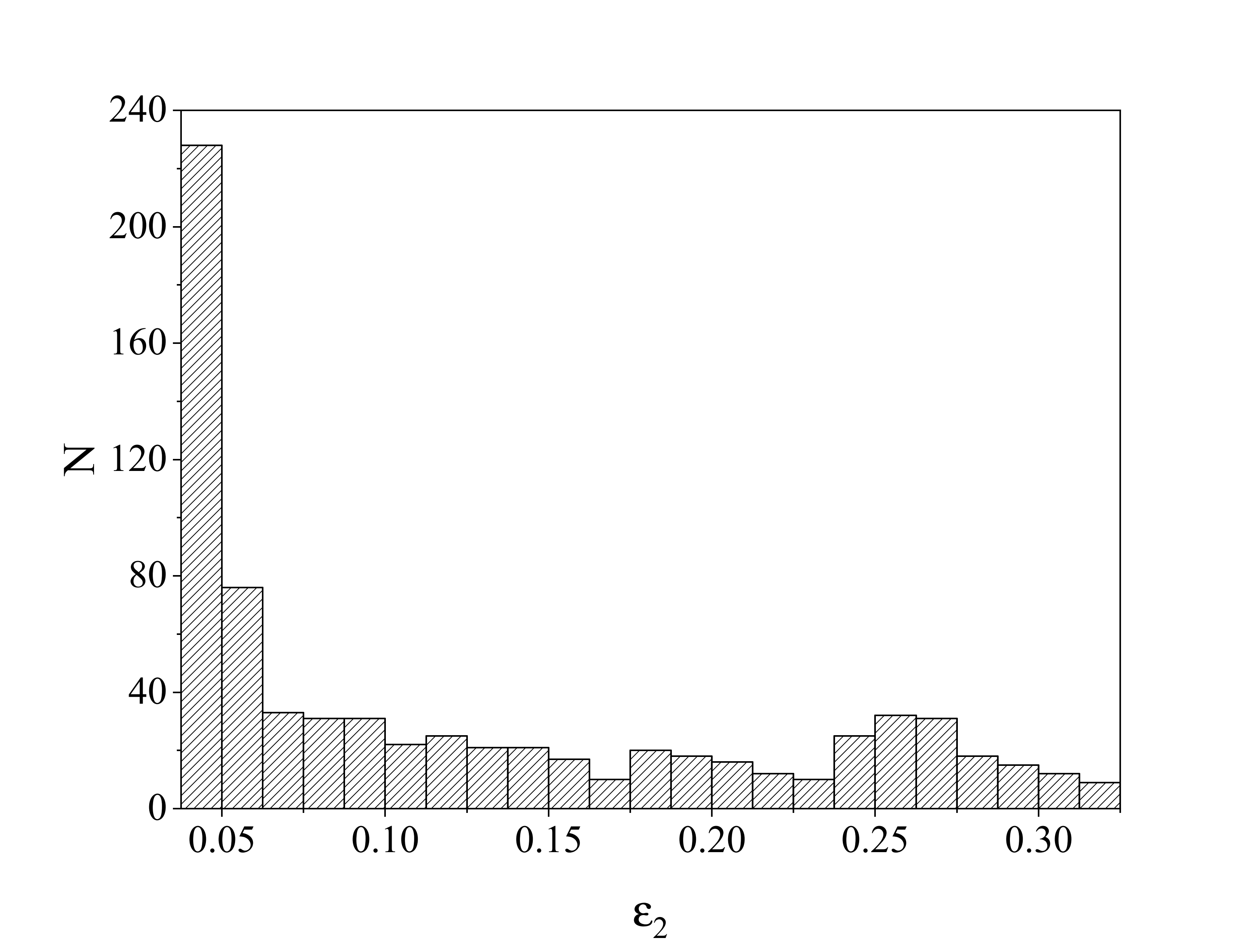}
\includegraphics[bb= 30 10 715 535, clip, width=0.5\linewidth]{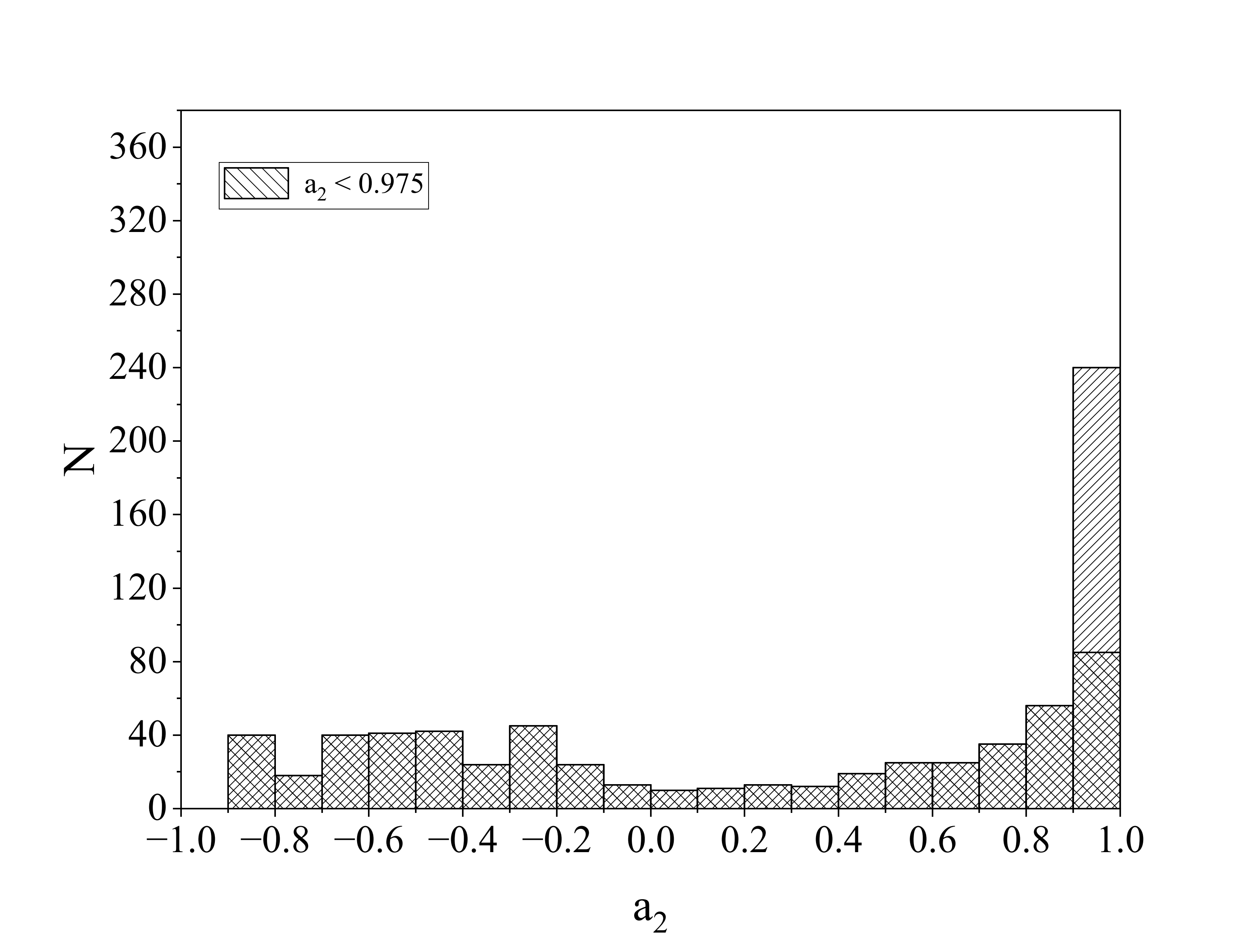}
\includegraphics[bb= 30 5 715 535, clip, width=0.5\linewidth]{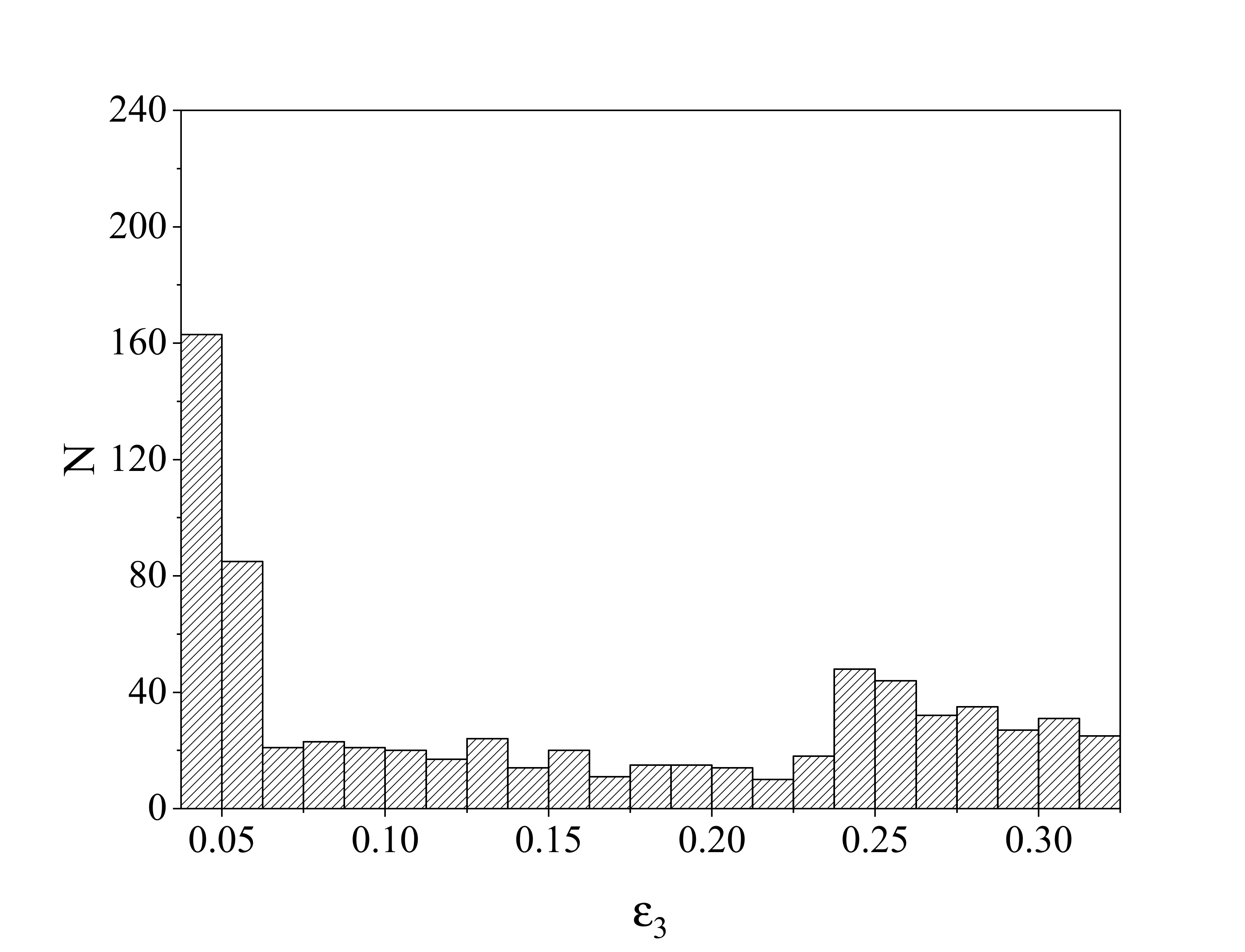}
\includegraphics[bb= 30 10 715 535, clip, width=0.5\linewidth]{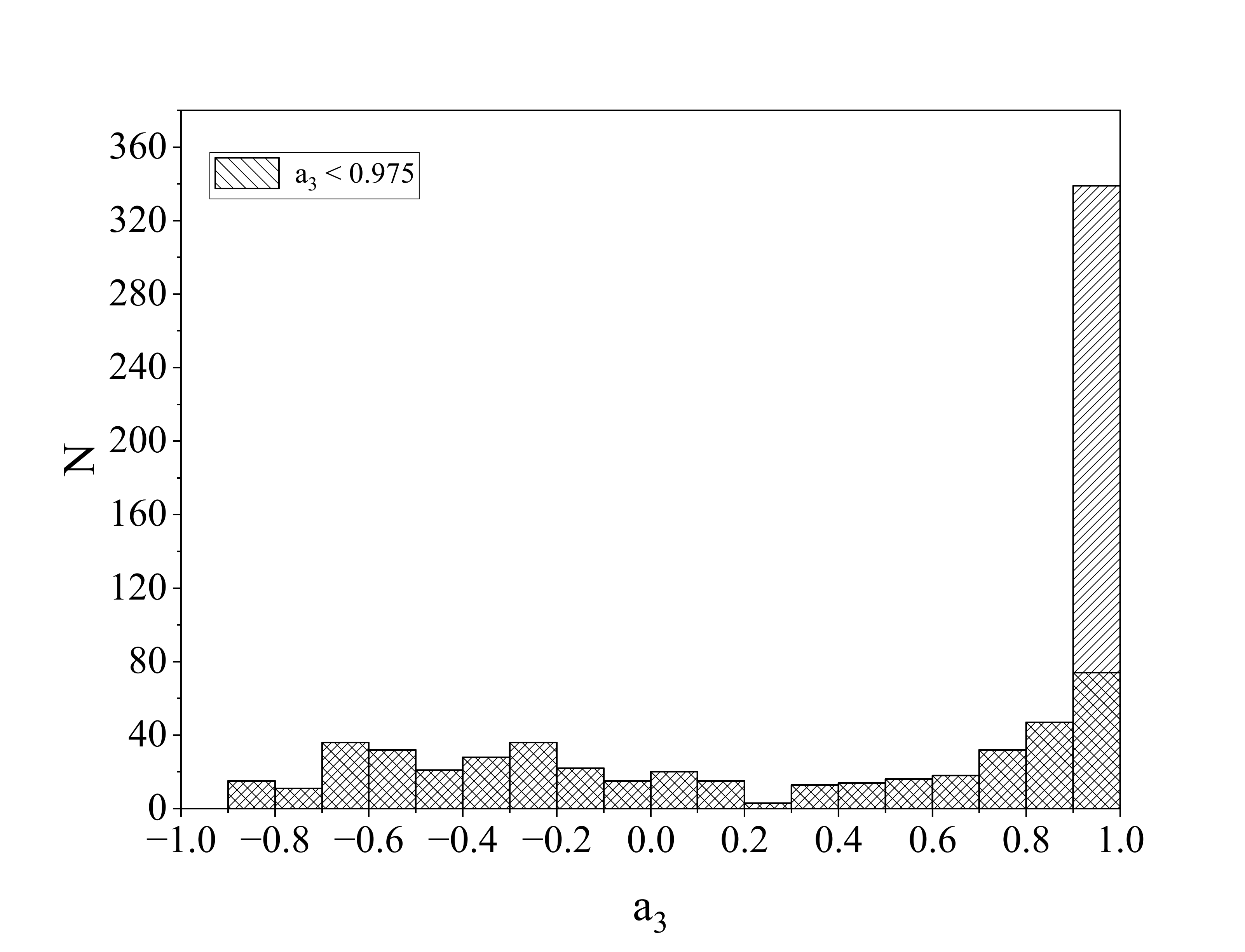}
\caption{Distribution of the estimated radiative efficiency and the spin values for all three models.
\label{fig:hist_a_eps}}
\end{figure}

\begin{figure}[ht!]
\includegraphics[bb= 30 5 715 535, clip, width=0.5\linewidth]{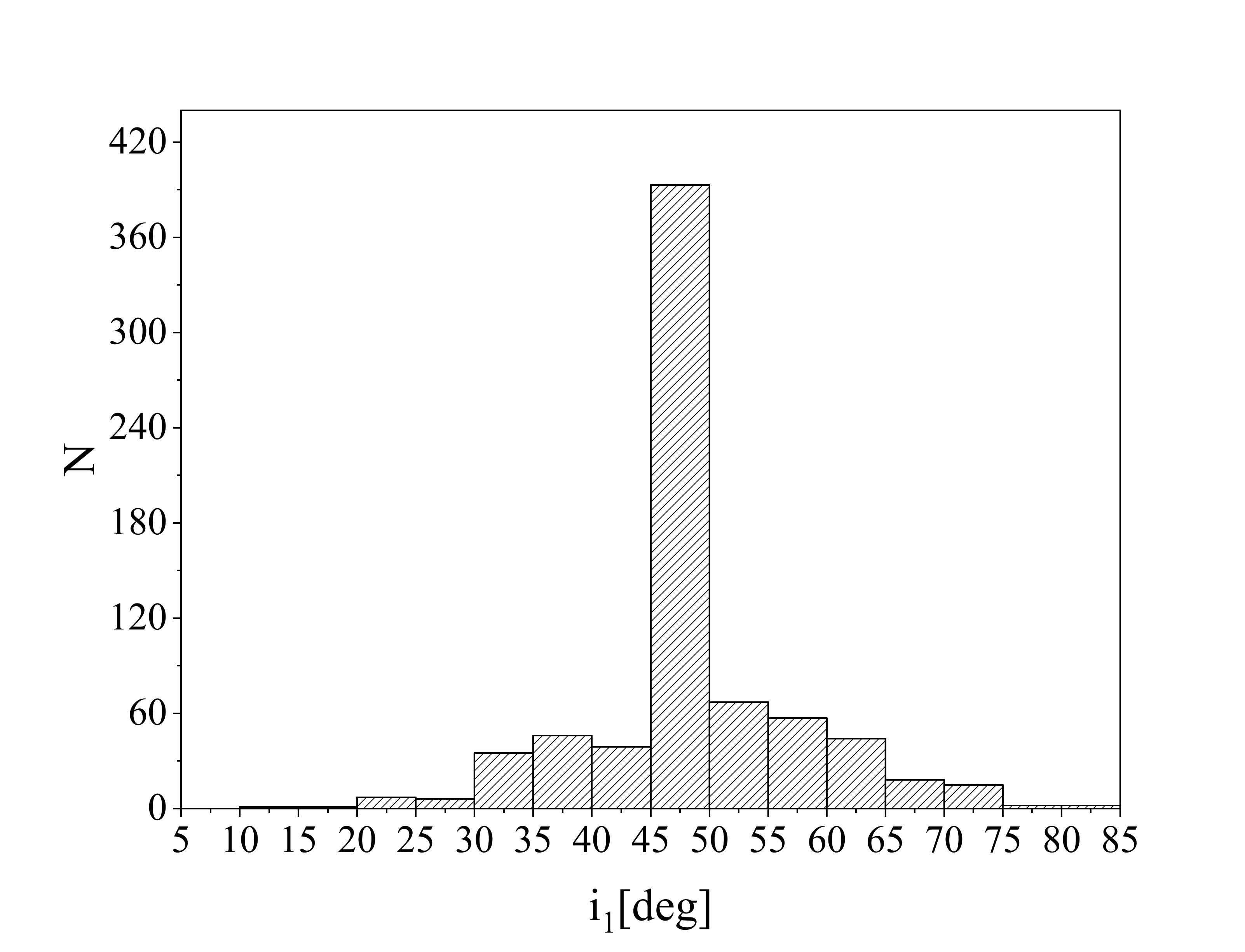}
\includegraphics[bb= 30 10 715 535, clip, width=0.5\linewidth]{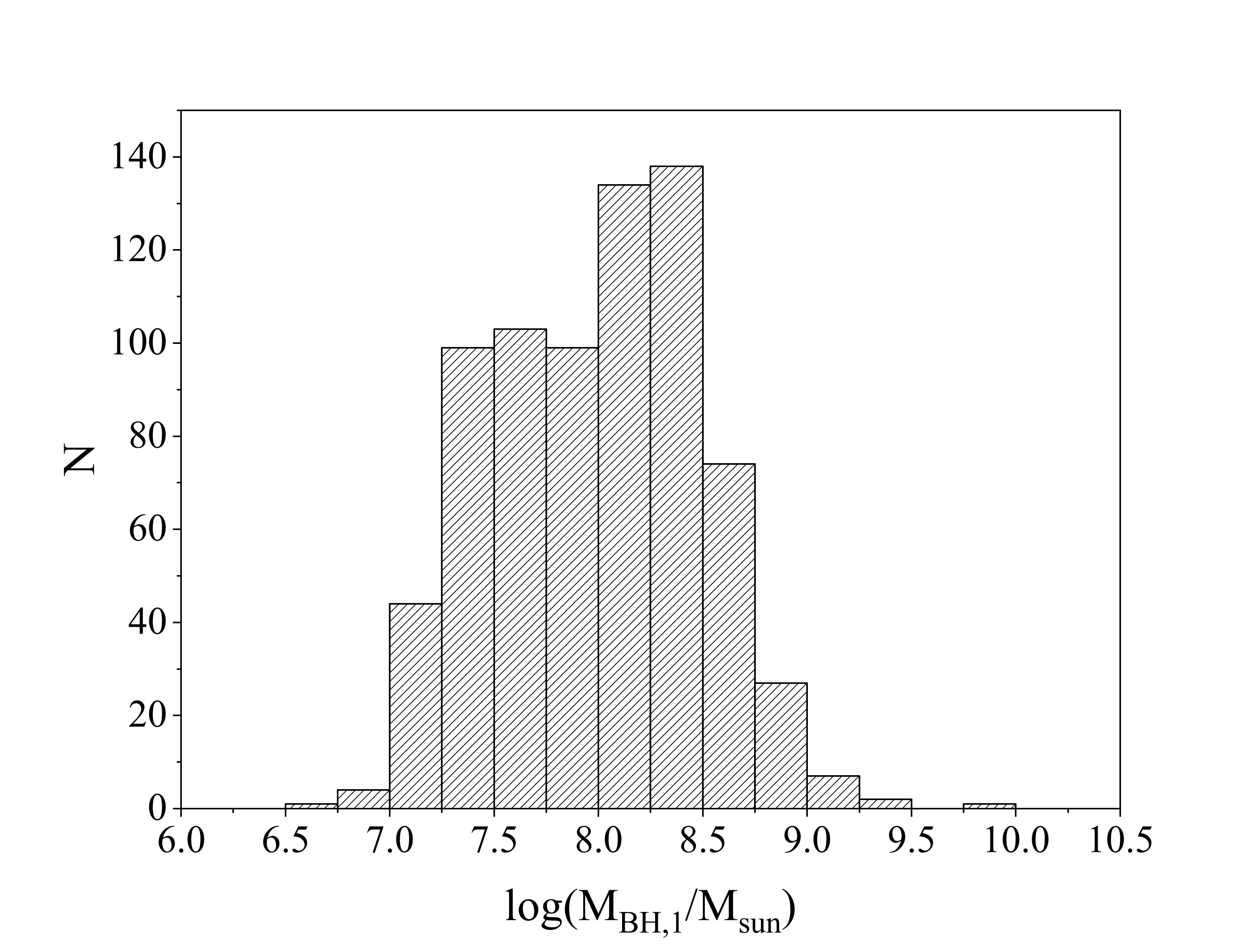}
\includegraphics[bb= 30 5 715 535, clip, width=0.5\linewidth]{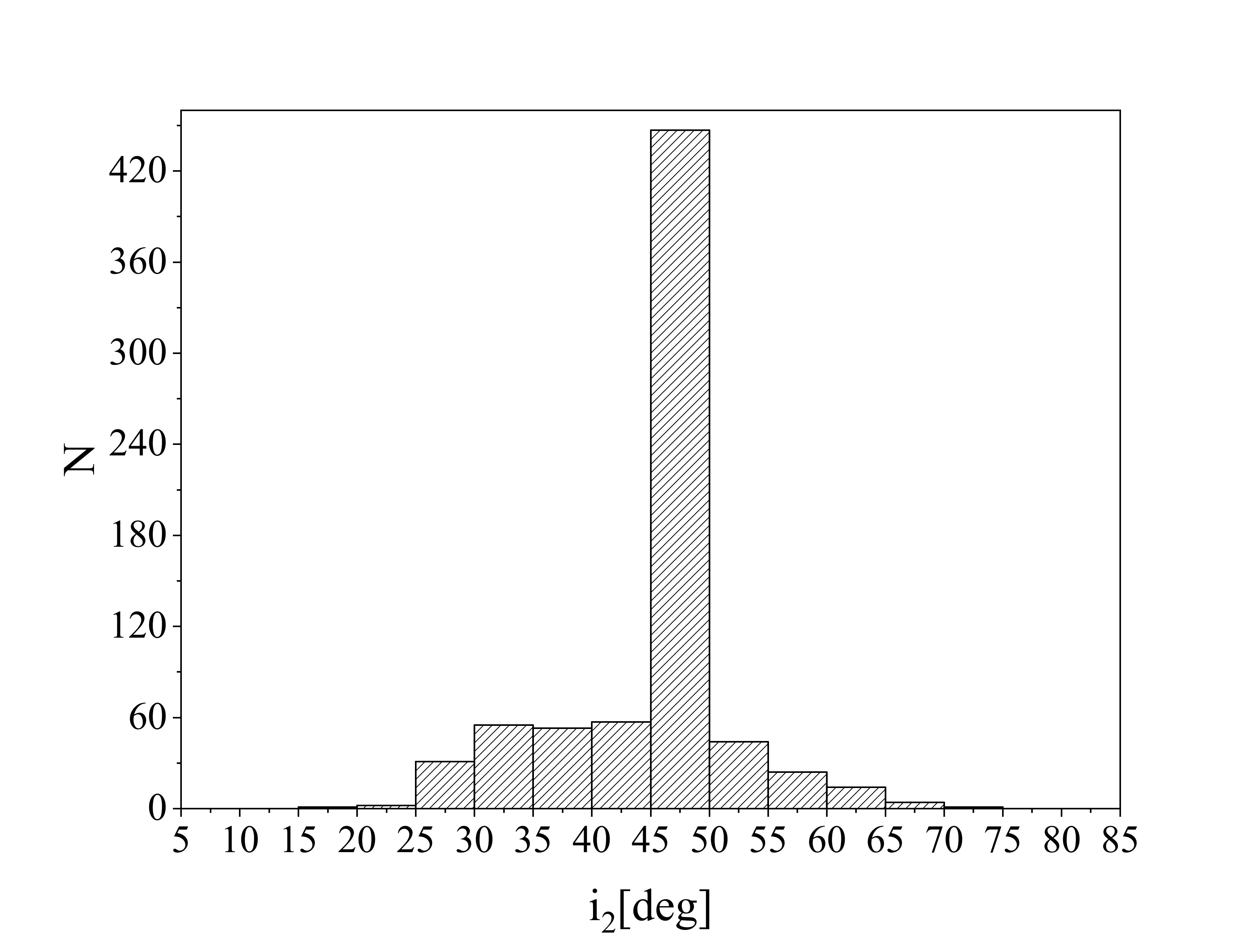}
\includegraphics[bb= 30 10 715 535, clip, width=0.5\linewidth]{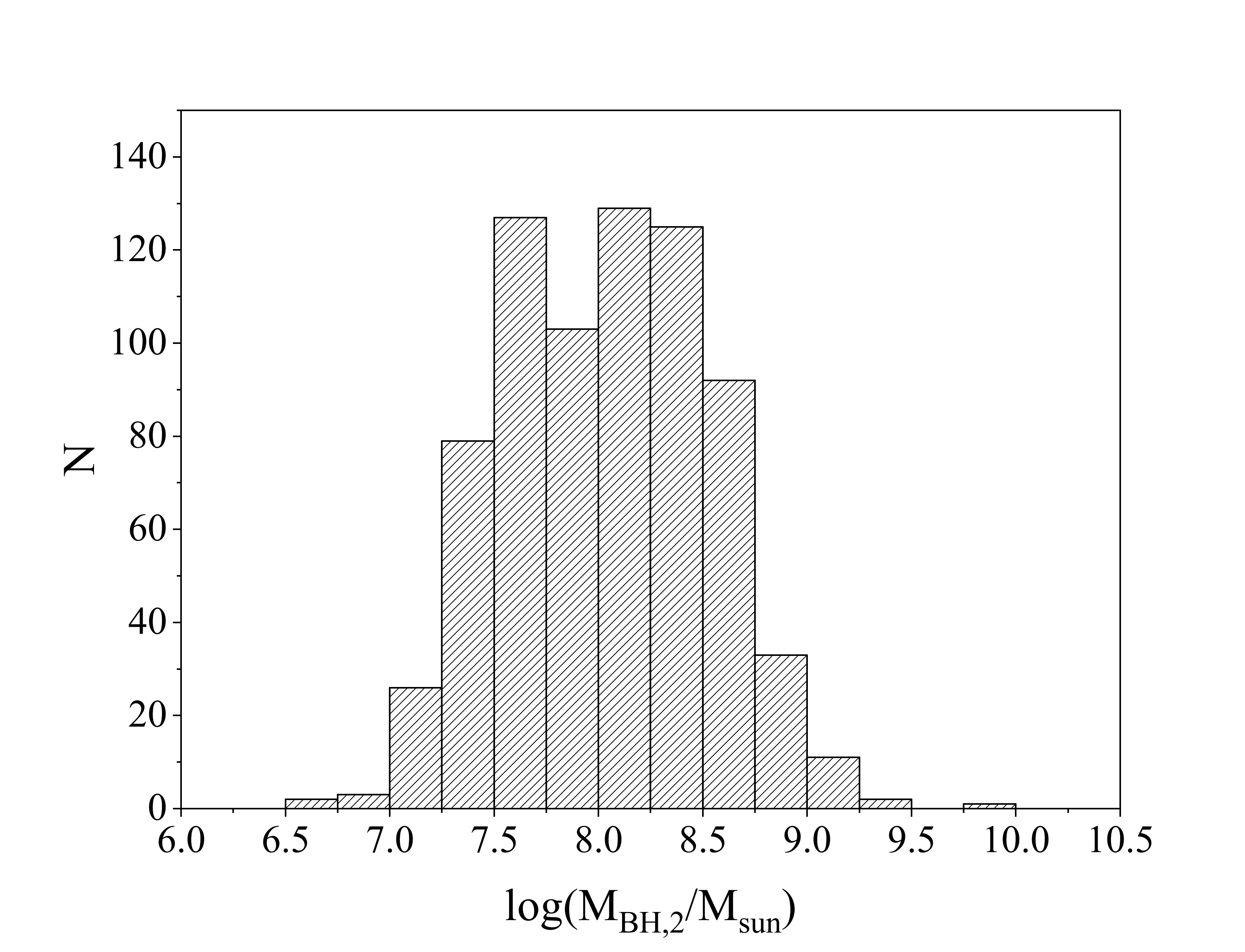}
\includegraphics[bb= 30 5 715 535, clip, width=0.5\linewidth]{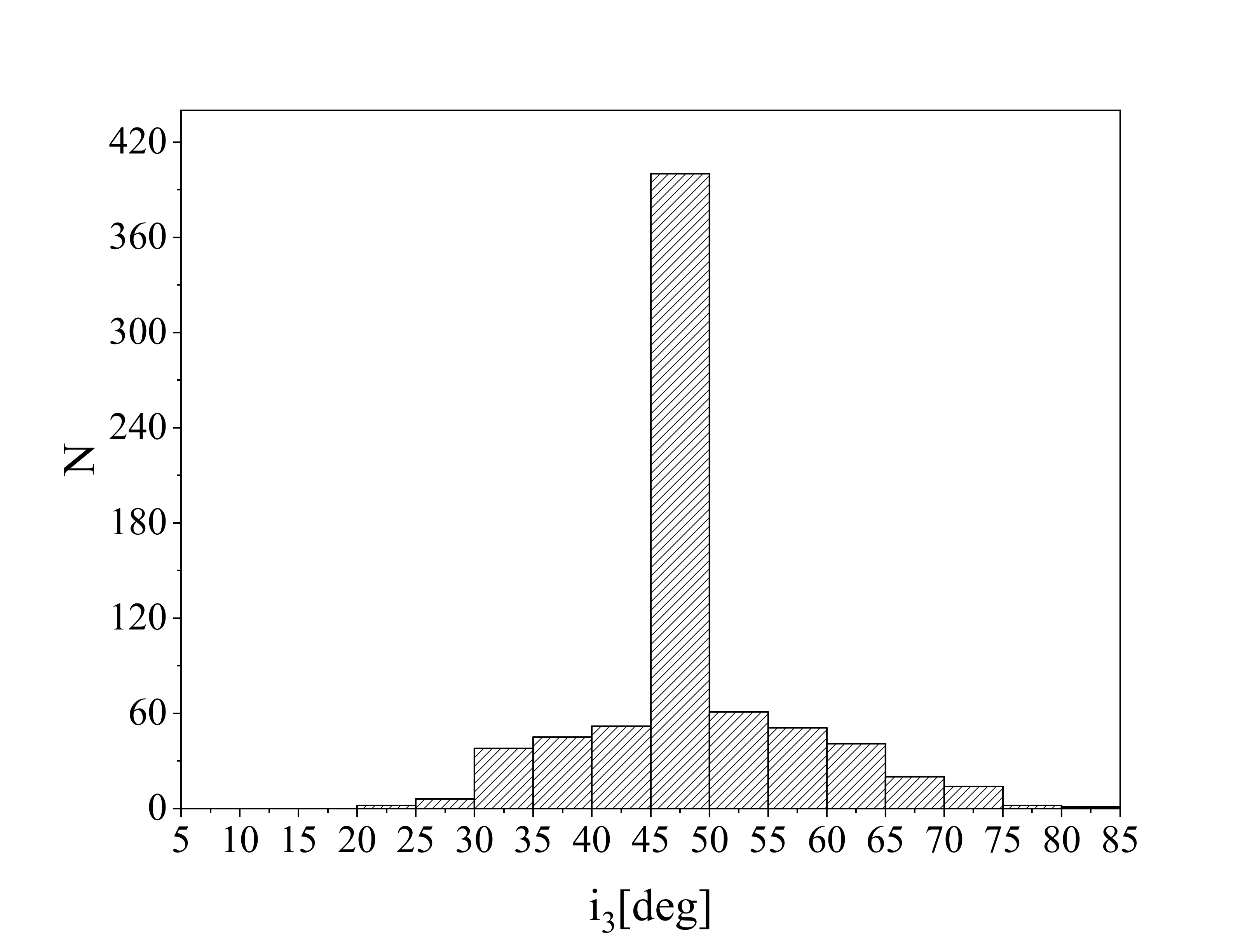}
\includegraphics[bb= 30 10 715 535, clip, width=0.5\linewidth]{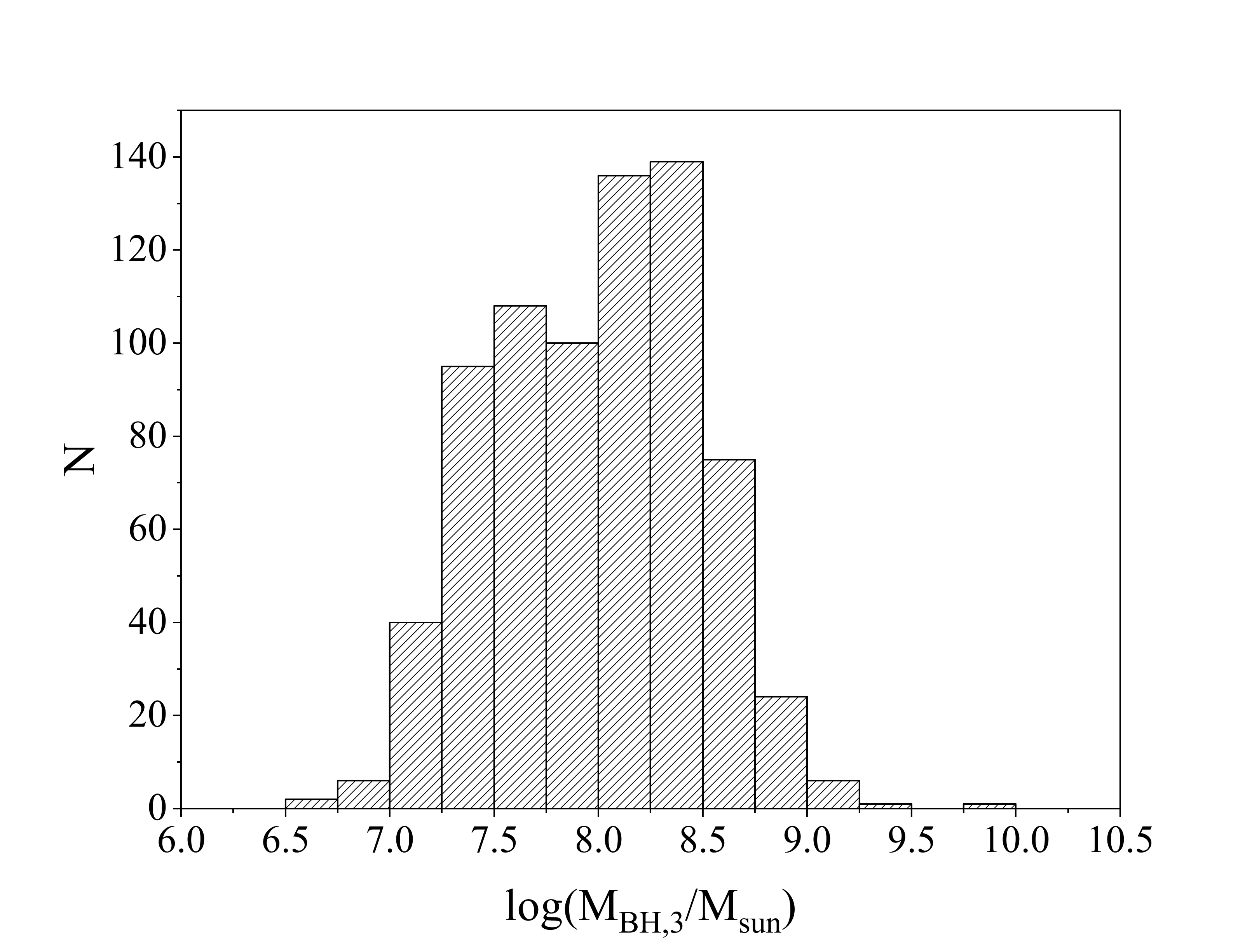}
\caption{Distribution of the estimated inclination angle and SMBH mass values for all three models.
\label{fig:hist_i_M}}
\end{figure}

\begin{figure}[ht!]
\includegraphics[bb= 30 5 715 535, clip, width=0.5\linewidth]{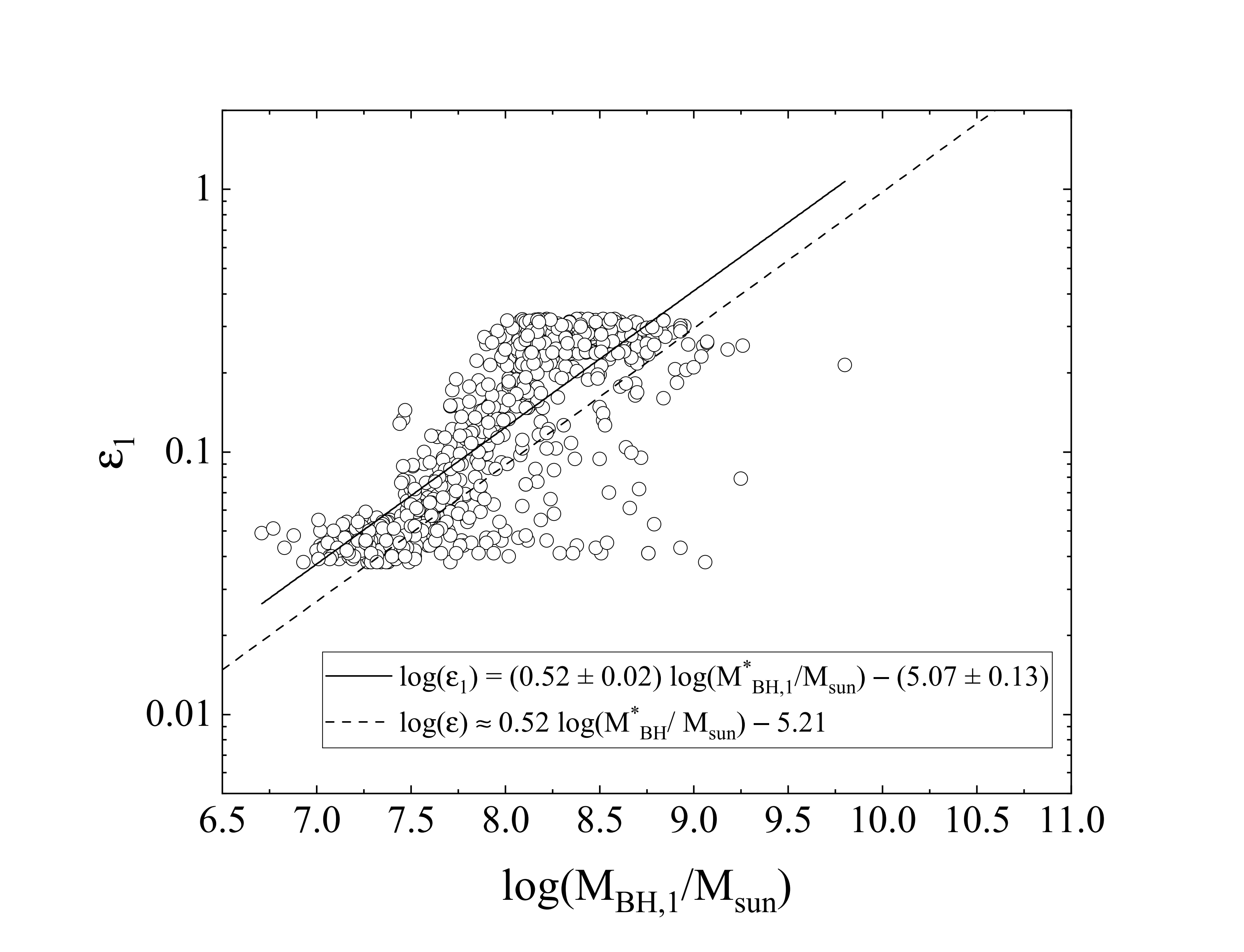}
\includegraphics[bb= 30 10 715 535, clip, width=0.5\linewidth]{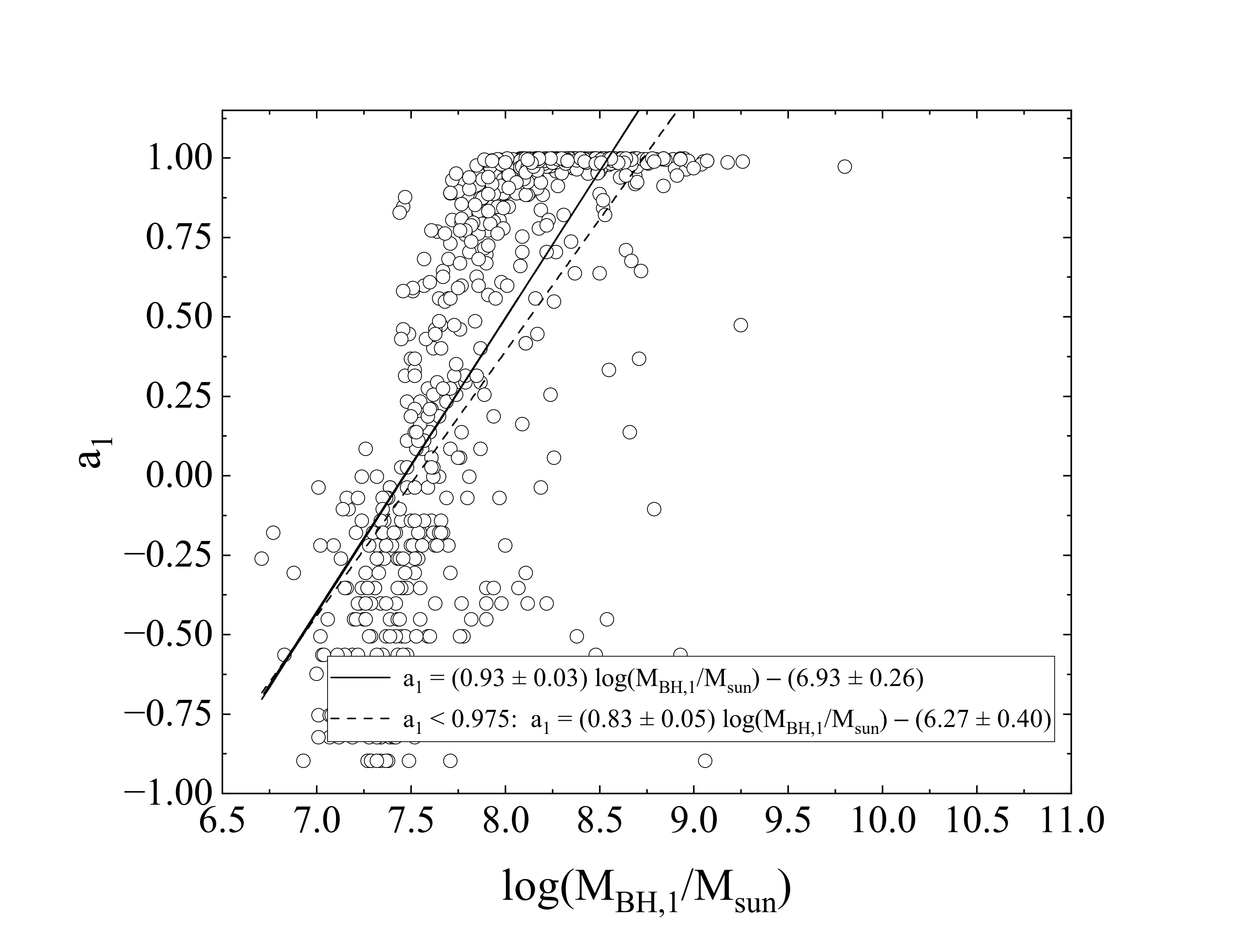}
\includegraphics[bb= 30 5 715 535, clip, width=0.5\linewidth]{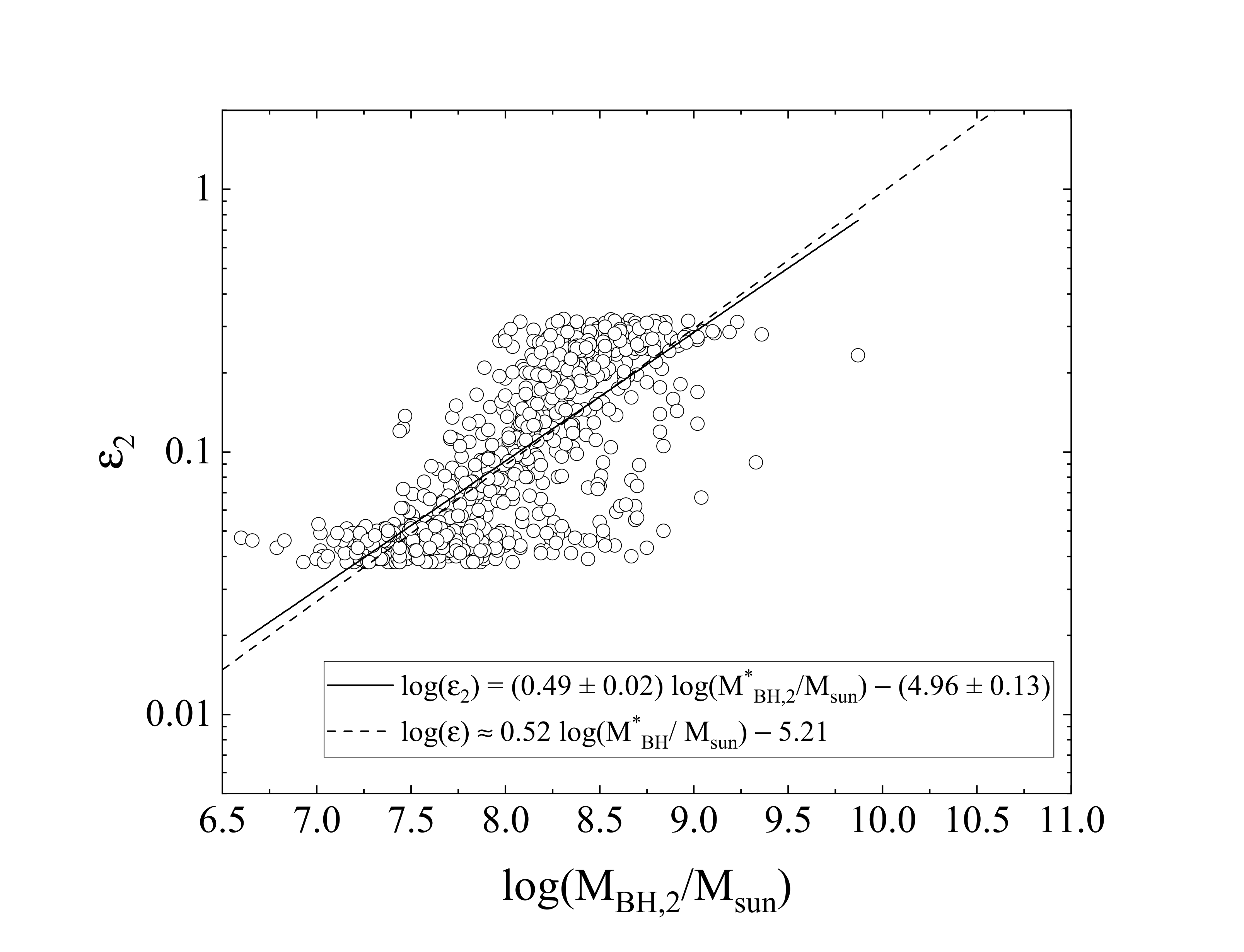}
\includegraphics[bb= 30 10 715 535, clip, width=0.5\linewidth]{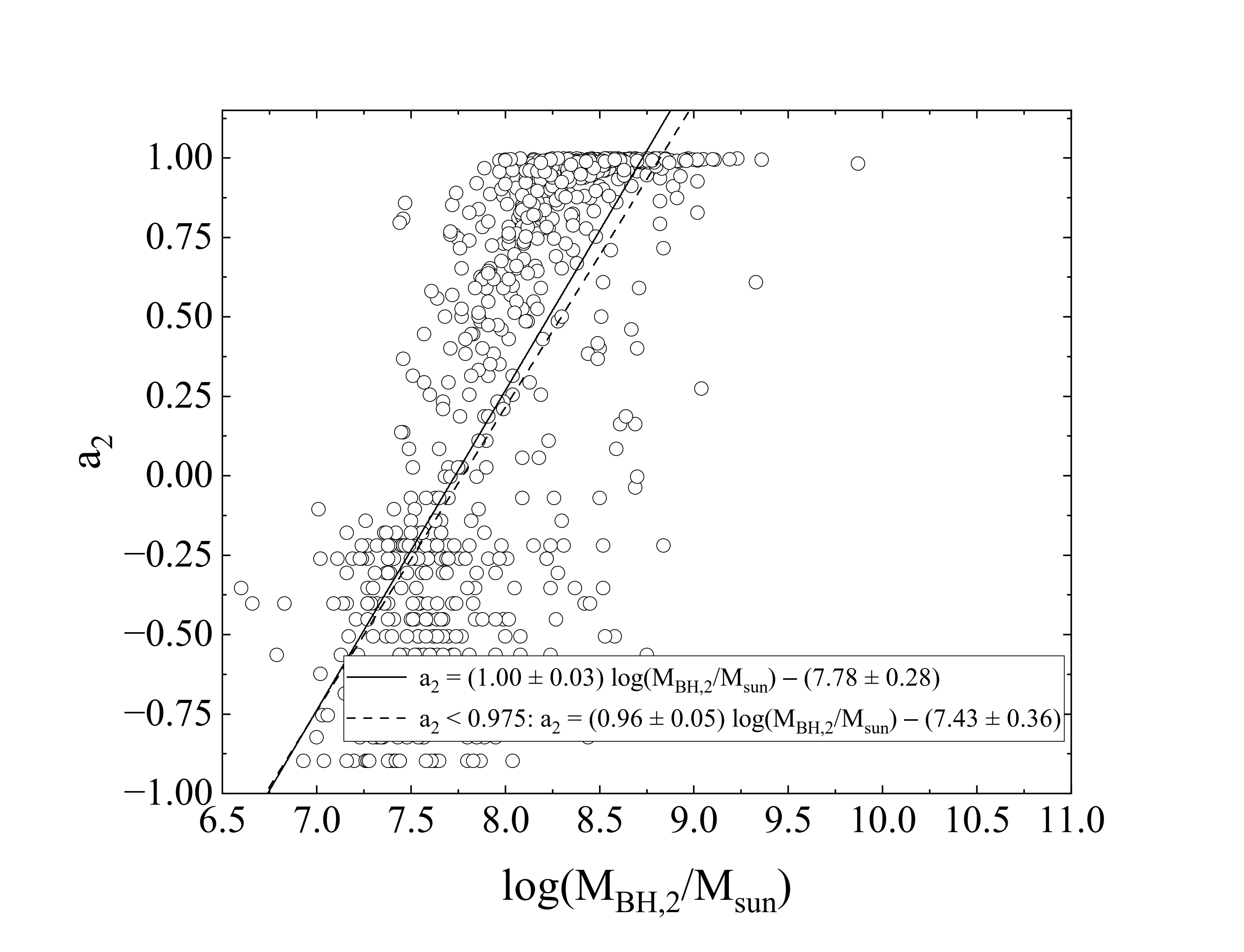}
\includegraphics[bb= 30 5 715 535, clip, width=0.5\linewidth]{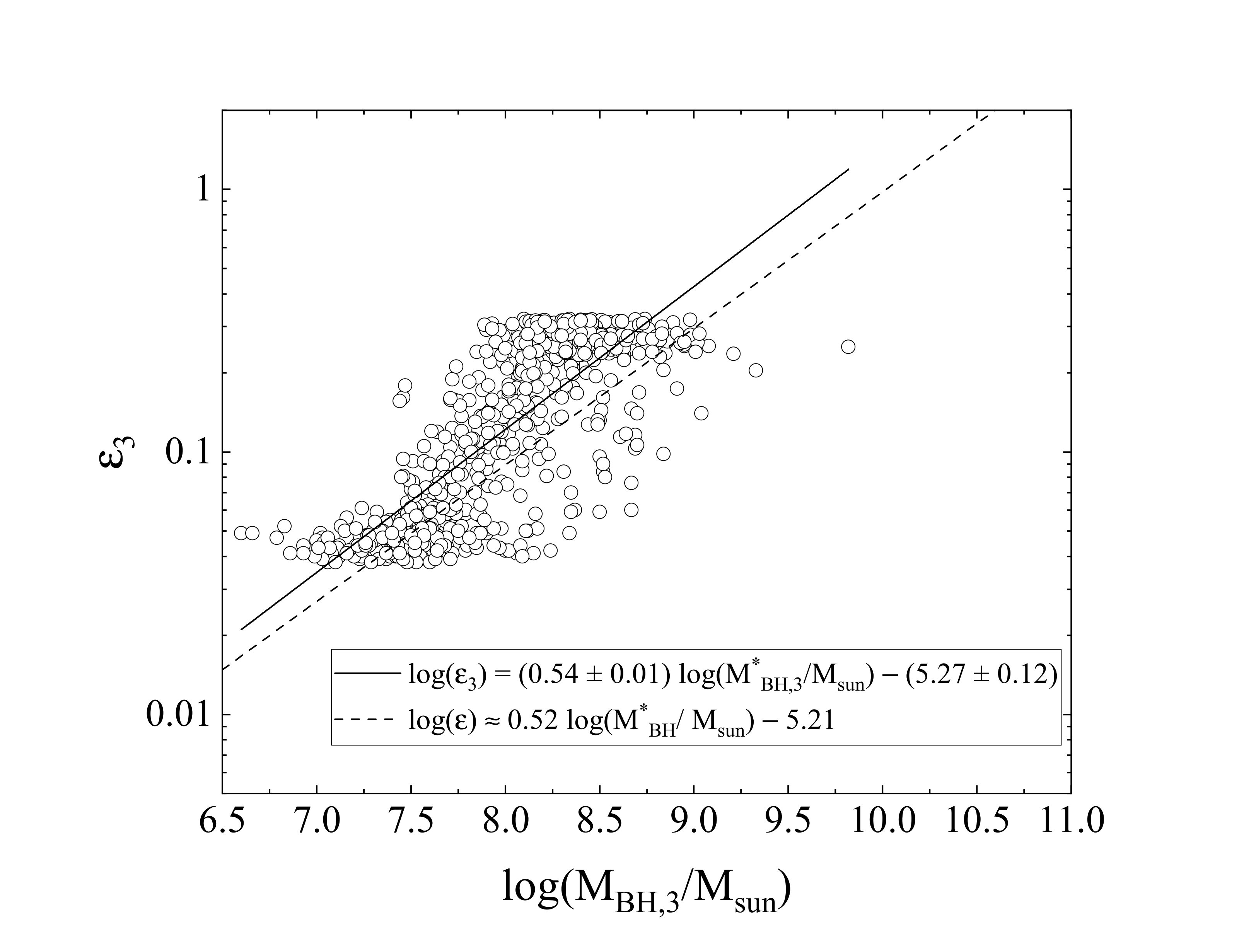}
\includegraphics[bb= 30 10 715 535, clip, width=0.5\linewidth]{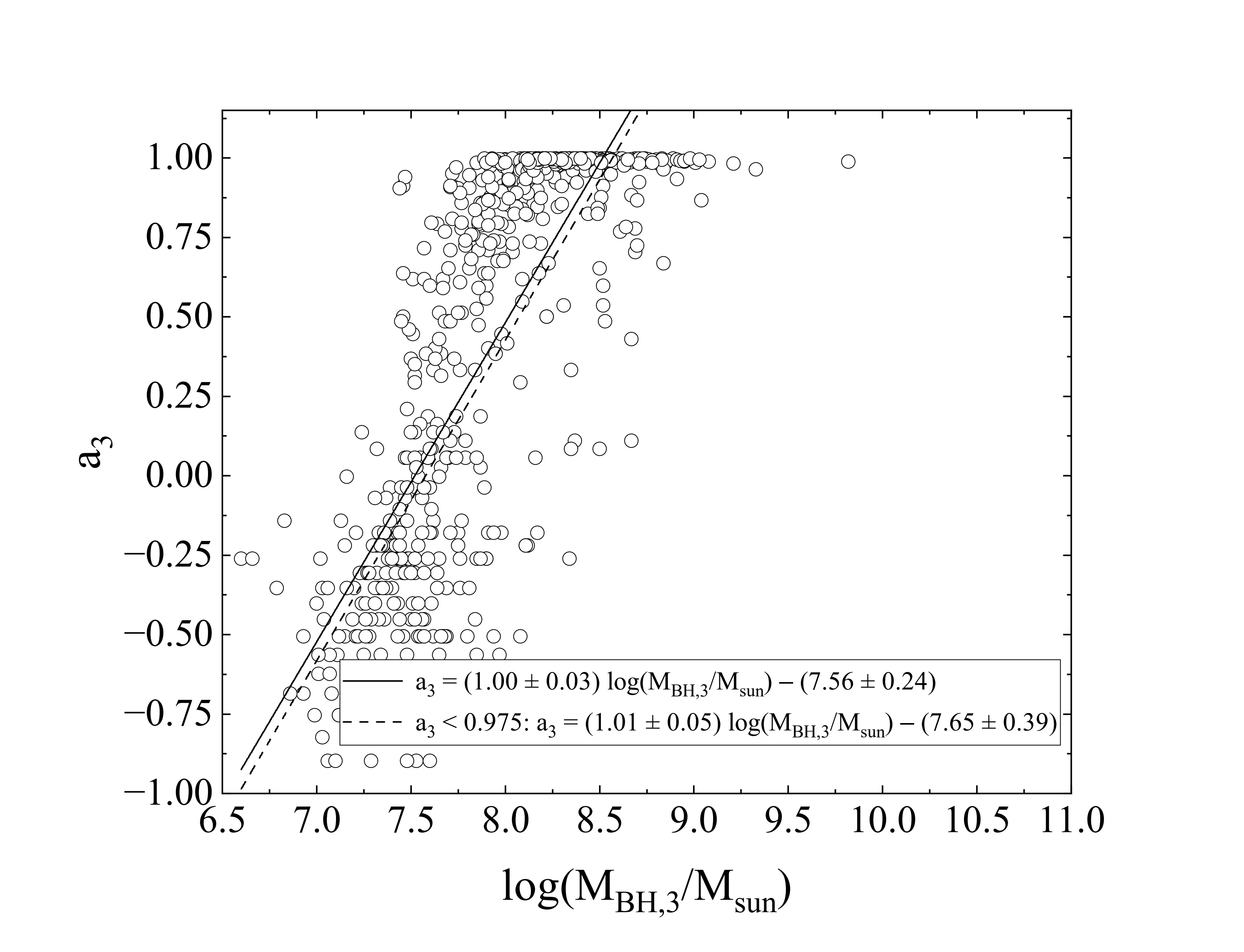}
\caption{Dependencies of the estimated radiative efficiency and the spin values on SMBH masses for all three models. The dashed line is the linear fit for the sample from \citet{davis11}.
\label{fig:a_MBH_eps_MBH}}
\end{figure}

We estimated the radiative efficiency, spins, inclination angles, and corresponding new SMBH masses and Eddington ratios for all 733 objects using three models. It should be noted that our results are largely statistically meaningful, we do not claim to precisely determine the spins of each specific object; this is a very complex task requiring extensive separate work. All this data is presented at Table \ref{tab:data}. The columns are: SDSS catalog designation, $z$ - cosmological redshift, $L_{\rm bol}$ - bolometric luminosity in $\log(L_{\rm bol}{\rm [erg/s]})$, $M_{\rm BH}$ - SMBH mass in $\log(M_{\rm BH} / M_\odot)$, $l_{\rm E}$ - Eddington ratio in $\log(l_{\rm E})$, $J$-$K_{\rm s}$ - difference between magnitudes in $J$ and $K_{\rm s}$ photometric bands, $r$-$K_{\rm s}$ - difference between magnitudes in $r$ and $K_{\rm s}$ photometric bands, $W1$-$W2$ - difference between magnitudes in W1 and W2 photometric bands. Estimated parameters for first model: $i_1$ - inclination angle (in degrees), $\varepsilon_1$ - radiative efficiency, $a_1$ - spin value, $M_{\rm BH,1}$ - SMBH mass, $l_{\rm E,1}$ - Eddington ratio. Next comes analogous data obtained for the second and third models.

Fig.\ref{fig:hist_a_eps} shows the distribution of the estimated radiative efficiency and the spin values for all three models. In general, the spin distribution looks typical for AGNs and quasars \citep{trakhtenbrot14,afanasiev18,piotrovich22,piotrovich24,piotrovich25b,daly19,reynolds21,azadi23} however, it is curious that all three models yielded negative spin values for as many as 190 objects. Such a large percentage of objects with retrograde rotation is quite unusual. The peak near 0 in the $\varepsilon$ distribution, which produces an unusually large number of objects with retrograde rotation, may be caused by an incorrect (underestimated) bolometric luminosity due to the reddened spectrum. If this is a real effect, it may indicate that when these objects are formed, they initially contain a large percentage of objects with retrograde rotation. From this we can assume that these are either very young objects or objects that formed as a result of mergers. Then, via disk accretion, the retrograde rotation of the SMBH gradually transforms into prograde rotation.

In Fig.\ref{fig:hist_i_M} one can see the distribution of the estimated inclination angle and SMBH mass values for all three models. Note that the precisely determining these angles is a very complex task, requiring either spectropolarimetric observations or methods such as echo mapping \citep{sha22,piotrovich23b} and our inclination angle values are only approximate estimates and we report these angles primarily to provide a more complete picture of how we perform our calculations. They are simply the angles at which our models yielded meaningful results. In this sense, the small fraction of values that strongly deviate from the initial default value of 45 degrees have the greatest physical meaning. This means, for example, that if our angle is significantly less than 45 degrees, then in reality it is also likely less than 45 degrees, and so on. The SMBH mass distributions have roughly the same shape as the distribution of the original red quasar sample, although the average masses are slightly smaller due to the fact that we use different inclination angles than the commonly used constant value of about 30-35 degrees.

The dependencies of the estimated spin on the cosmological redshift and the bolometric luminosity do not show a correlation between the parameters (Pearson correlation coefficient $\leq 0.05$).

Fig.\ref{fig:a_MBH_eps_MBH} shows the dependencies of the estimated radiative efficiency and the spin values on SMBH masses for all three models. There are strong correlation between parameters. Pearson correlation coefficients for $\varepsilon(M_{\rm BH})$ relations for three models are 0.77, 0.76 and 0.80 and for $a(M_{\rm BH})$ relations they are 0.73, 0.73 and 0.78 respectively. Linear fitting gives us:
\begin{equation}
  \begin{aligned}
    &\log(\varepsilon_1) = (0.52 \pm 0.02) \log(M_{\rm BH,1} / M_\odot)) - (5.07 \pm 0.13),\\
    &\log(\varepsilon_2) = (0.49 \pm 0.02) \log(M_{\rm BH,2} / M_\odot)) - (4.96 \pm 0.13),\\
    &\log(\varepsilon_3) = (0.54 \pm 0.01) \log(M_{\rm BH,3} / M_\odot)) - (5.27 \pm 0.12),
  \end{aligned}
\end{equation}
\begin{equation}
  \begin{aligned}
    &a_1 = (0.93 \pm 0.03) \log(M_{\rm BH,1} / M_\odot)) - (6.93 \pm 0.26),\\
    &a_2 = (1.00 \pm 0.03) \log(M_{\rm BH,2} / M_\odot)) - (7.78 \pm 0.28),\\
    &a_3 = (1.00 \pm 0.03) \log(M_{\rm BH,3} / M_\odot)) - (7.56 \pm 0.24).
  \end{aligned}
\end{equation}

The dashed line is the linear fit for the sample from \citet{davis11}: $\log(\varepsilon) \approx 0.52 \log(M_{\rm BH,1} / M_\odot)) - 5.21$. One can see that our linear fittings are very close to this fitting. Thus, we can conclude that our objects in this regard are similar in their statistical properties to AGN in general.

To exclude possible saturation effects at high spin values, we additionally examined a sample of objects with $a < 0.975$ for all three models. It can be seen (Fig.\ref{fig:hist_a_eps}) that for such samples, the distribution becomes much more uniform. The resulting approximations are quite close to the original ones. Thus, it can be concluded that in this case the saturation effect does not have a significant impact on the result.

\begin{figure}[ht!]
\centering
\includegraphics[bb= 30 5 715 535, clip, width=0.7\linewidth]{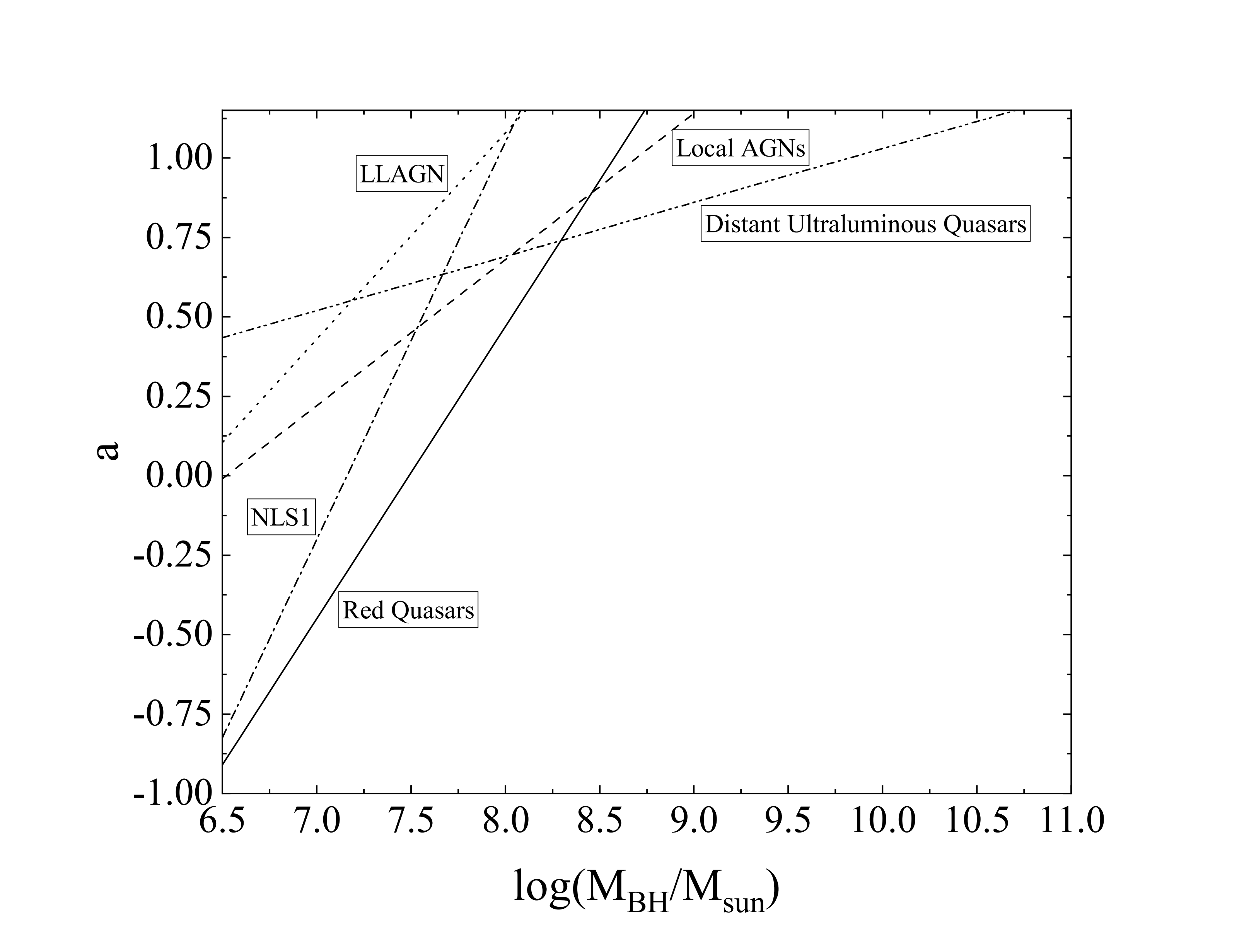}
\caption{The spin--mass dependence linear fits for different types of objects.
\label{fig:fits}}
\end{figure}

\begin{table}[ht!]
\begin{center}
\caption[]{The spin--mass dependence linear fits for different types of objects. LLAGN is Low Luminosity AGN.}
\begin{tabular}{lll}
\hline\noalign{\smallskip}
{\bf Object type}             & {\bf Linear fit}                                                  & {\bf Reference}\\
\hline\noalign{\smallskip}
Distant Ultraluminous Quasars &  $a = (0.17 \pm 0.05) \log(M_{\rm BH}/M_\odot) - (0.67 \pm 0.49)$ &  \citet{piotrovich25}\\
Local AGNs                    &  $a = (0.46 \pm 0.09) \log(M_{\rm BH}/M_\odot) - (3.00 \pm 0.71)$ &  \citet{piotrovich22}\\
LLAGN                         &  $a = (0.65 \pm 0.11) \log(M_{\rm BH}/M_\odot) - (4.12 \pm 0.82)$ &  \citet{piotrovich25b}\\
Red Quasars                   &  $a = (0.93 \pm 0.03) \log(M_{\rm BH}/M_\odot) - (6.93 \pm 0.26)$ &  Present work \\
NLS1                          &  $a = (1.25 \pm 0.05) \log(M_{\rm BH}/M_\odot) - (8.95 \pm 0.35)$ &  \citet{piotrovich23}\\
\noalign{\smallskip}\hline
\end{tabular} \label{tab:fits}
\end{center}
\end{table}

The spin--mass slope of 0.9-1 is somewhat larger than that obtained in the previous study (0.7-0.8) \citep{piotrovich24}, which is most likely due to the fact that the new sample contains significantly more objects and more different types of objects. This slope is significantly larger than the similar slope, for example, for local AGNs (0.3-0.45) \citep{piotrovich22}. Of all the objects we have studied previously in terms of spin--mass slope red quasars are closest to Narrow Line Seyfert Type 1 galaxies (NLS1) ($\sim 1.25$) \citep{piotrovich23} (See Fig.\ref{fig:fits} and Table \ref{tab:fits}). This confirms the conclusion from our previous work \citep{piotrovich24} that red quasars are likely to contain both Seyferts and NLS1. In general, from the fact that the spin of red quasars increases rapidly with the SMBH mass, we can conclude that the main mechanism of SMBH mass growth in these objects is disk accretion.

\section{Conclusions}

Using ''color cut'' method we obtained from SDSS DR16 catalog 733 red quasar candidates, which amounted to approximately 4\% of the objects for which the catalog contained all the data necessary for our calculation (17926 objects).

Then we estimated the radiative efficiency, spins, inclination angles, and corresponding new SMBH masses for all 733 objects using three theoretical models. In general, obtained spin distributions looked typical for AGNs and quasars however, it is curious that all three models yielded negative spin values for as many as 190 objects. Such a large percentage of objects with retrograde rotation is quite unusual. It may indicate that when these objects are formed, they initially contain a large percentage of objects with retrograde rotation, which may show that these are either very young objects or objects that formed as a result of mergers. The dependencies of the estimated spin values on SMBH masses show characteristic trend with a slope of about 0.9-1.0 which allows us to assume that red quasars are likely to contain both Seyferts and NLS1, and that the main mechanism of SMBH mass growth in these objects is disk accretion. It should be clarified that all of these results are highly model-dependent, and, furthermore, the observational data on which they are based have significant uncertainties. However, obtaining more accurate results requires extensive and comprehensive observations (spectropolarimetry and/or echo mapping) of each individual object, which should be the subject of separate future studies.

\begin{acknowledgements}
The authors are grateful to the reviewer for very useful comments.
\end{acknowledgements}

\newpage
\appendix

\setcounter{table}{0}
\renewcommand{\thetable}{A\arabic{table}}

{\tiny
\setlength\tabcolsep{1pt}
\begin{longtable}{lcccccccccccccccccccccc}
\caption{Results of our estimations for three models.\label{tab:data}}\\
\hline\noalign{\smallskip}
SDSS name & $z$ & $L_{\rm bol}$ & $M_{\rm BH}$ & $l_{\rm E}$ & $J$-$K_{\rm s}$ & $r$-$K_{\rm s}$ & $W1$-$W2$ & $i_1$ & $\varepsilon_1$ & $a_1$ & $M_{\rm BH,1}$ & $l_{\rm E,1}$ & $i_2$ & $\varepsilon_2$ & $a_2$ & $M_{\rm BH,2}$ & $l_{\rm E,2}$ & $i_3$ & $\varepsilon_3$ & $a_3$ & $M_{\rm BH,3}$ & $l_{\rm E,3}$ \\
\hline\noalign{\smallskip}
000236.24-002724.7 & 0.291000 & 45.06 &  8.51 & -1.57 & 1.74 & 4.36 & 0.88 & 45 & 0.239 &  0.984 &  8.10 & -1.15 & 45 & 0.147 &  0.884 &  8.10 & -1.15 & 45 & 0.230 &  0.980 &  8.10 & -1.15\\
000445.71+002245.0 & 0.496237 & 45.42 &  8.32 & -1.01 & 1.77 & 4.18 & 0.90 & 45 & 0.102 &  0.696 &  7.90 & -0.60 & 45 & 0.060 &  0.110 &  7.90 & -0.60 & 45 & 0.090 &  0.598 &  7.90 & -0.60\\
000713.21+021816.5 & 0.297854 & 44.79 &  7.29 & -0.61 & 2.65 & 4.77 & 1.17 & 30 & 0.053 & -0.106 &  7.17 & -0.49 & 30 & 0.044 & -0.506 &  7.17 & -0.49 & 35 & 0.038 & -0.898 &  7.06 & -0.37\\
000805.26+265317.3 & 0.197850 & 44.74 &  7.60 & -0.98 & 1.73 & 4.02 & 1.06 & 45 & 0.043 & -0.564 &  7.19 & -0.56 & 40 & 0.049 & -0.262 &  7.27 & -0.64 & 45 & 0.045 & -0.452 &  7.19 & -0.56\\
000828.11-000210.2 & 0.258000 & 44.73 &  8.14 & -1.53 & 1.74 & 4.17 & 0.87 & 45 & 0.150 &  0.890 &  7.73 & -1.11 & 45 & 0.112 &  0.758 &  7.73 & -1.11 & 45 & 0.158 &  0.908 &  7.73 & -1.11\\
000855.68+053457.6 & 0.241000 & 44.92 &  8.40 & -1.60 & 1.85 & 4.19 & 1.06 & 45 & 0.221 &  0.976 &  7.99 & -1.18 & 45 & 0.145 &  0.880 &  7.99 & -1.18 & 45 & 0.219 &  0.974 &  7.99 & -1.18\\
\hline
\end{longtable}}

\bibliographystyle{raa}
\bibliography{mybibfile}

\end{document}